\documentclass[sn-mathphys-num, twocolumn]{revtex4-2}

\usepackage{amssymb,amsmath}
\usepackage{graphicx}
\usepackage{dcolumn}
\usepackage{bm}
\usepackage{color}
\usepackage[a4paper,margin=1in]{geometry}
\usepackage[ruled]{algorithm2e}
\usepackage[colorlinks=true, allcolors=blue]{hyperref}
\usepackage{enumerate}
\usepackage{booktabs}
\usepackage{xcolor}
\usepackage{comment}
\usepackage{dcolumn}%
\usepackage{titlesec}

\titlespacing*{\section}{0pt}{10pt}{0pt}
\titleformat{\section}[hang]{\normalfont\bfseries}{\thesection}{1ex}{}

\newcommand{\beginMainText}{%
        \setcounter{figure}{0}
        \renewcommand{\thefigure}{\textbf{\arabic{figure}}}%
        \renewcommand{\figurename}{\textbf{Fig.}}%
}

\newcommand{\beginExtendedData}{%
        \setcounter{figure}{0}
        \renewcommand{\thefigure}{\textbf{\arabic{figure}}}%
        \renewcommand{\figurename}{\textbf{Extended Data Fig.}}%
}
\newcommand{\beginMethodology}{%
        \setcounter{equation}{0}
        \renewcommand{\theequation}{M\arabic{equation}}%
        \setcounter{section}{0}
        \renewcommand{\thesection}{\arabic{section}}%
        \setcounter{subsection}{0}
        \renewcommand{\thesubsection}{\thesection.\arabic{subsection}}
     }

\newcommand{\beginSI}{%
        \setcounter{table}{0}
        \renewcommand{\thetable}{\textbf{S\arabic{table}}}%
        \renewcommand{\tablename}{\textbf{Table}}%
        \setcounter{figure}{0}
        \renewcommand{\thefigure}{\textbf{S\arabic{figure}}}%
        \renewcommand{\figurename}{\textbf{Fig.}}%
        \setcounter{section}{0}
        \renewcommand{\thesection}{\arabic{section}}%
        \setcounter{subsection}{0}
        \renewcommand{\thesubsection}{\thesection.\arabic{subsection}}
        \setcounter{subsubsection}{0}
        \renewcommand{\thesubsubsection}{\thesubsection.\arabic{subsubsection}}
        \setcounter{equation}{0}
        \renewcommand{\theequation}{S\arabic{equation}}%
     }

\newcommand{\rv}[1]{\begingroup #1\endgroup}

\begin{document}

\title{Metamaterials that learn to change shape}

\author{Yao Du$^{1}$}
\author{Ryan van Mastrigt$^{1,2,3}$}
\author{Jonas Veenstra$^{1}$}
\author{Corentin Coulais$^{1,}$}
\email{c.j.m.coulais@uva.nl}
\affiliation{
$^{1}$ Institute of Physics, University of Amsterdam, Science Park 904, 1098 XH, Amsterdam, The Netherlands\\
$^{2}$ AMOLF, Science Park 104, 1098 XG Amsterdam, The Netherlands\\
$^{3}$ Gulliver UMR CNRS 7083, ESPCI Paris, PSL University, 10 rue Vauquelin, 75005 Paris, France}
\date{\today}

\begin{abstract}
Learning to change shape is a fundamental strategy of adaptation and evolution of living organisms, from cells to tissues and animals. Human-made materials can also exhibit advanced shape morphing capabilities, but lack the ability to learn. Here, we build metamaterials that can learn complex shape-changing responses using a contrastive learning scheme. By being shown examples of the target shape changes, our metamaterials are able to learn those shape changes by progressively updating internal learning degrees of freedom---the local stiffnesses. Unlike traditional materials that are designed once and for all, our metamaterials have the ability to forget and learn new shape changes in sequence, to learn multiple shape changes that break reciprocity, and to learn multistable shape changes, which in turn allows them to perform reflex gripping actions and locomotion. Our findings establish metamaterials as an exciting platform for physical learning, which in turn opens avenues for the use of physical learning to design adaptive materials and robots. 
\end{abstract}


\maketitle
\beginMainText

\section*{Introduction}
One of the distinctive functionalities of living materials, such as biological polymers, cells, tissues, and living organisms, is the ability to change shape. A frontier of material science is to create synthetic materials that emulate these shape-changing capabilities. Over the past years, metamaterials have emerged as a prominent platform to do so all the way from the micron~\cite{smart_magnetically_2024, liu_electronically_2025} to the centimeter~\cite{coulais_combinatorial_2016, overvelde_rational_2017, kim_printing_2018, siefert_bio-inspired_2019, choi_programming_2019, zareei_harnessing_2020, jin_kirigamiinspired_2020, van_manen_4d_2021, hwang_shape_2022, gao_pneumatic_2023, meeussen_multistable_2023} and meter scale~\cite{melancon_multistable_2021, meeussen_multistable_2023, stein-montalvo_kirigami-inspired_2024, li_adaptive_2024}. 
These metamaterials may impact a range of applications from biomedicine~\cite{van_manen_4d_2021, smart_magnetically_2024}, robotics~\cite{hwang_shape_2022, smart_magnetically_2024, li_adaptive_2024, baines_robots_2024, liu_electronically_2025} and architecture~\cite{melancon_multistable_2021, adrover_deployable_2015, stein-montalvo_kirigami-inspired_2024, li_adaptive_2024}. Yet, these shape-morphing metamaterials miss a crucial property that is prevalent in living materials: the ability to adapt their shape-changing response to changing conditions and to learn by modifying their components locally after fabrication~\cite{bastien_unifying_2013, tala_pseudomonas_2019, noselli_swimming_2019, kramar_encoding_2021}. 

Here, inspired by recent developments in physical learning~\cite{williamChemicalBoltzmannMachines2017, sternSupervisedLearningPhysical2020a, stern_supervised_2021, dillavou_demonstration_2022, stern_learning_2023, falk_learning_2023, patil_self-learning_2023, ArinzePRE2023, altman_experimental_2024, dillavou_machine_2024, mandal_learning_2024, falk_temporal_2025}, we create metamaterials that learn to change shape. The general framework of physical learning aims to emulate nature's ability to learn in physical systems by systematically adjusting a system's internal parameters---the so-called learning degrees of freedom---using a predefined local learning rule, thereby evolving the system towards a desired response. \rv{Unlike shape memory materials, which can be trained to memorize one or two shapes~\cite{jani2014review}, our metamaterials can be trained instead to change shape in response to a mechanical input.}
Trained under a local supervised physical learning scheme, our metamaterials can learn, forget, relearn new shape changes on demand, and even learn multiple target shapes simultaneously. Notably, our learning scheme generalizes to energy non-conserving cases, viz., with nonreciprocity~\cite{fruchart_odd_2023, mandal_learning_2024, veenstra_non-reciprocal_2024, veenstra_adaptive_2025}, and nonlinear cases, viz., with multistability~\cite{melancon_multistable_2021, meeussen_multistable_2023, veenstra_non-reciprocal_2024}. Taken together, these learned nonreciprocal and multistable shape changes endow our metamaterials with robotic functionalities such as reflex gripping and locomotion. Our study demonstrates that metamaterials are a powerful platform for physical learning and paves the way toward adaptive materials and robots.

\begin{figure*}[htbp!]
    \centering
    \includegraphics[width=1\textwidth]{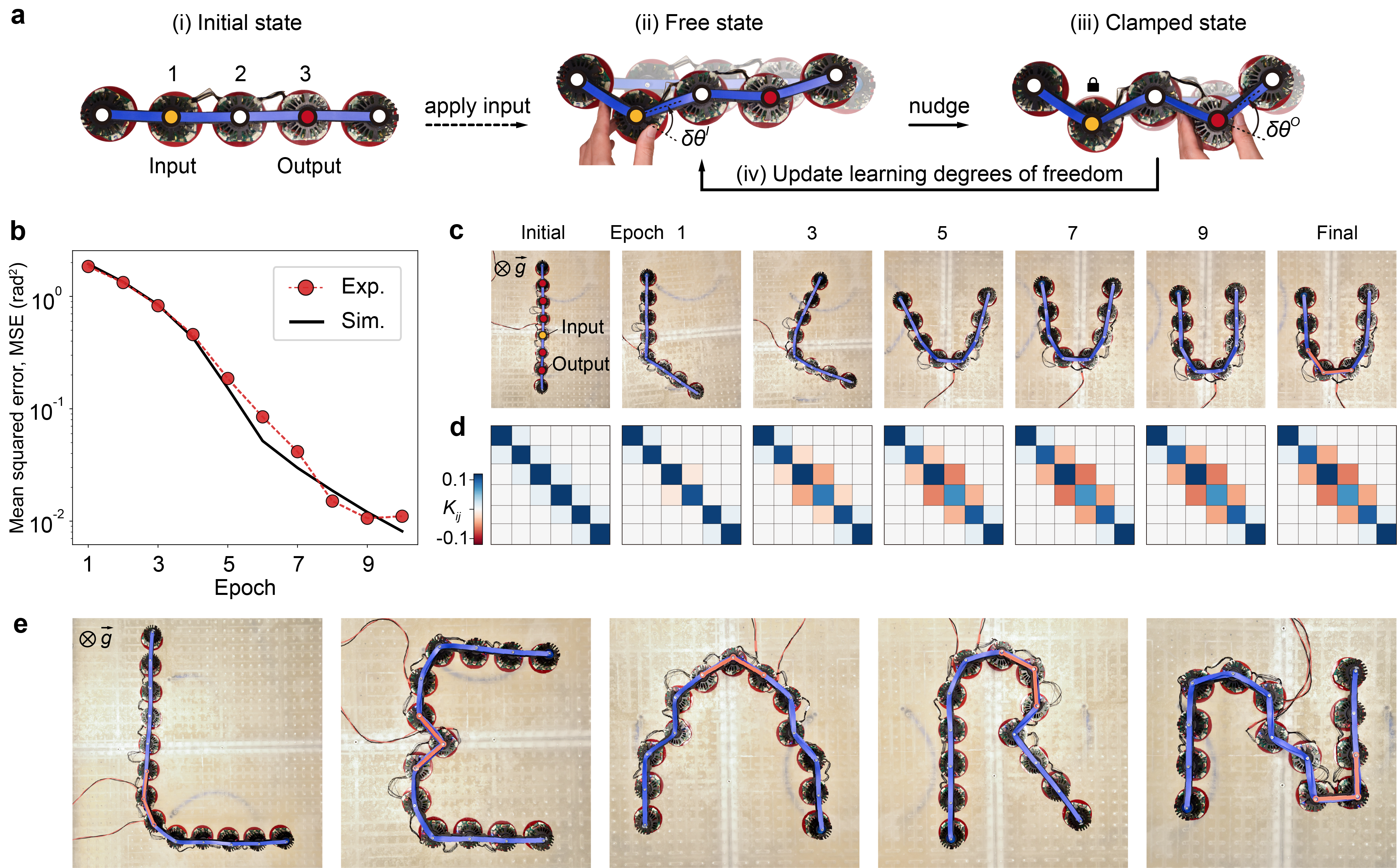}
    \caption{\textbf{Contrastive learning for shape-changing metamaterials.} 
    {\bf a,} Contrastive learning scheme. In the free state, the system is deformed from its initial equilibrium state by the input angle $\delta\theta^{I}$, whereas in the clamped state, both the input $\delta\theta^{I}$ and the desired output $\delta\theta^{O}$ are kept fixed. During learning,  steps (ii-iv) are repeated while the learning degrees of freedom are updated according to the contrastive learning rule until a predetermined number of epochs is reached. 
    {\bf b,} The \rv{mean squared error (MSE) curves} in simulation (solid line) and experiment (red dots) where a $N=6$ robotic chain is trained to morph into a U-shape. Here, the learning rate is $\gamma=0.01$. \rv{See the simulation protocol in the Methodology.}
    {\bf c,} Equilibrium configurations of each epoch in the free state. Note that the two edge units are not actuated. 
    {\bf d,} The stiffness matrix $K$ during learning. \rv{$K_{ij}$ refers the entry on $i^{\text{th}}$ row and the $j^{\text{th}}$ column in the stiffness matrix. Note that $K$ is a tridiagonal matrix since only the nearest-neighbor interaction is involved here.} The initial parameters are $k_{i}^{o}=0.1$, $k_{i}^{p}=0.01$ and $k_{i}^{a}=0$. Note $k^{e}$ is a constant and thus not shown. 
    {\bf e,} A metamaterial with $N=11$ is sequentially trained to form the word ``LEARN". See Extended Data Fig.~\ref{figE:continue learning} for the corresponding MSE curves. The red linkage applies the input angular deflection.}
    \label{fig:Learning demenstration}
\end{figure*}

\section*{Experimental setup}
We construct a robotic metamaterial made from $N$ units consisting of motorized hinges able to exert a torque. The units are connected by an elastic skeleton (Fig.~\ref{fig:Learning demenstration}a and Extended Data Fig.~\ref{figE:robotic unit}, see \rv{Methodology} for details). Additionally, each unit has a microcontroller that measures its own angular deflections $\delta\theta_{i}$ and exchanges information with its nearest neighbors, stores memory of their past deformations, and applies programmable torques via a local feedback loop. These capabilities allow us to adjust the local stiffness of units as we see fit and to implement a torque on each unit $i$ as
\begin{equation}
\label{eq:tau_i}
\begin{split}
\tau_{i}=&-\left(k_{i}^{o}+k^{e}\right)\delta\theta_{i}
         -\left(k_{i-1}^{p}+k_{i-1}^{a}\right)\delta\theta_{i-1}\\
         &-\left(k_{i}^{p}-k_{i}^{a}\right)\delta\theta_{i+1},    
\end{split}
\end{equation}
where $k_{i}^{o}$, $k_{i}^{p}$ and $k_{i}^{a}$ are the on-site stiffness, the passive (symmetric) neighbor stiffness, and the active (anti-symmetric) neighbor stiffness. These parameters can be manipulated via the local active feedback loop. $k^{e}$ is the stiffness of the elastic skeleton and is fixed. We conduct our experiments on a low-friction air table on which the metamaterial can freely move. 
\rv{We apply external deformations by manually fastening some of the units with screws.}
Doing so generates a torque through the elastic skeleton and active control (Eq.~\eqref{eq:tau_i}) so that the metamaterial evolves towards a new mechanical equilibrium. In what follows, we aim to control this mechanical equilibrium as a function of the imposed external deformations. We will first consider reciprocal interactions ($k_{i}^{a}=0$) and then generalize our findings to path-dependent non-reciprocal scenarios ($k_{i}^{a}\neq0$). 

\section*{Contrastive learning scheme} 
To control the shape changes of our metamaterial, we apply a form of physical learning called contrastive learning~\cite{movellan_contrastive_1991, williamChemicalBoltzmannMachines2017, stern_supervised_2021}. Contrastive learning uses the difference between two states of mechanical equilibrium, the free and clamped states, to define a local learning rule. In the free state, only input deformations are imposed. In the clamped state, both input and desired output deformations are imposed simultaneously. The goal is to adjust the learning degrees of freedom to achieve the desired output deformations when imposing a predefined input deformation.

In our metamaterial, the angular deflections $\delta\theta_{i}$ are the so-called physical degrees of freedom: variables that follow from the physical laws governing the system. The tunable stiffnesses $k_{i}^{o}$, $k_{i}^{p}$ and $k_{i}^{a}$ are the learning degrees of freedom: parameters that can be tuned and, crucially, influence the resulting physical degrees of freedom. We aim to find an optimal set of stiffnesses that achieves the desired angular deflection $\delta\theta^{O}$ for the output units by applying a predefined angular deflection $\delta\theta^{I}$ to the input units. Consequently, our metamaterials can morph into a given shape with certain input angular deflections.

To find these stiffnesses that correspond to a desired shape change, we train our metamaterials following a supervised learning protocol (Fig.~\ref{fig:Learning demenstration}a):
\begin{enumerate}
    \item[(i)] Initialization. We set the straight chain as the reference configuration, i.e., $\delta\theta_{i}=0$ for all $i$. We determine the initial $k_{i}^{o}$ and $k_{i}^{p}$ but set $k_{i}^{a}=0$, resulting in a symmetric stiffness matrix~$K$.
    \item[(ii)] We apply fixed input angles $\delta\theta^{I}_i$.
    The current equilibrium configuration---the free state---is memorized in the microcontroller of each unit.
    \item[(iii)] While keeping the input units fixed, we clamp the output units to the desired angle $\delta\theta_{i}^{O}$ and store the new equilibrium configuration---the clamped state.
    \item[(iv)] The units compute new stiffnesses using the following local learning rule (Eqs.~\eqref{eq:kodot} and~\eqref{eq:kpdot}) and then update the parameters by a gradient descent step.
\end{enumerate}

The learning protocol consists of repeating steps (ii-iv) for multiple epochs. The local learning rule follows from the gradient of the difference between the function $\psi(\{\delta\theta\}, \{k\})$ evaluated in the free (F) and clamped (C) states:
\begin{equation}
\label{eq:dkdt}
    \frac{\mathrm{d} k_{i}}{\mathrm{d}t}=-\gamma\frac{\partial}{\partial k_{i}}\left(\psi^{C}-\psi^{F}\right), 
\end{equation}
where $\gamma$ is the learning rate and the superscript denotes in which state the function is evaluated. If the metamaterial is passive, i.e., $k_{i}^{a}=0$, its forces derive from a scalar potential. For such a system, $\psi$ is the elastic energy:
\begin{equation}
\label{eq:psi_p}
    \psi=\dfrac{1}{2}\sum_{i=1}^{N}\left(k_{i}^{o}+k^{e}\right)\left(\delta\theta_{i}\right)^{2}+\sum_{i=1}^{N-1}k_{i}^{p}\delta\theta_{i}\delta\theta_{i+1},
\end{equation}
where the first term represents the on-site energy of each unit and the second term captures the interaction energy between neighboring units. We then substitute Eq.~\eqref{eq:psi_p} into Eq.~\eqref{eq:dkdt} to obtain an explicit learning rule for the passive metamaterial:
\begin{equation}
\label{eq:kodot}
    \frac{\mathrm{d} k_{i}^{o}}{\mathrm{d}t} = -\dfrac{\gamma}{2}\left[\left(\delta\theta_{i}^{C}\right)^{2}-\left(\delta\theta_{i}^{F}\right)^{2}\right],
\end{equation}
\begin{equation}
\label{eq:kpdot}
    \frac{\mathrm{d} k_{i}^{p}}{\mathrm{d}t} = -\gamma\left(\delta\theta_{i}^{C}\delta\theta_{i+1}^{C}-\delta\theta_{i}^{F}\delta\theta_{i+1}^{F}\right),
\end{equation}
where $\delta\theta_{i}^{F}$ and $\delta\theta_{i}^{C}$ are the angular deflections of the $i^\text{th}$ unit in the free and clamped states respectively. Note that this learning rule is local because it involves only the angles of unit $i$ and neighboring unit $i+1$. Employing such a local learning rule over a central one as used in, e.g., back-propagation, requires only local flow of information and is therefore scalable. 

\section*{Learning to change shape}
We first demonstrate the learning procedure with a metamaterial with $N=6$ units. Our metamaterial learns to form the letter ``U" starting from a straight chain when applying an input of $\delta\theta_{3}=\pi/3$. Here, all other units are outputs. In the free state, we apply only the input in each epoch. In the clamped state, we nudge the chain to the desired shape by fastening the output units in addition to the input units. Using the angular deflections in these two states, each robotic unit calculates $\mathrm{d} k_{i}/\mathrm{d}t$ (Eqs.~\eqref{eq:kodot} and \eqref{eq:kpdot}) and subsequently updates all $k_{i}$. 

During the entire learning procedure (Video~S2), the mean square error, $\mathrm{MSE}=\textstyle\sum_{i}^{}(\delta\theta_{i}^{F}-\delta\theta_{i}^{O})^{2}/N_{O}$, gradually decreases and reaches values below 1\% after just 10 iterations in both the simulation and the experiment (Fig.~\ref{fig:Learning demenstration}b). Here, $N_{O}$ is the number of output units. As expected, this coincides with the metamaterial progressively converging to the desired ``U" shape in the free state (Fig.~\ref{fig:Learning demenstration}c) and an evolving stiffness matrix (Fig.~\ref{fig:Learning demenstration}d). 

To further challenge our metamaterial, we use a longer chain of $N=11$ and learn to form all the letters of the word ``LEARN'' sequentially as shown in Fig.~\ref{fig:Learning demenstration}e and Video~S2. Crucially, our metamaterial can forget the previous shape change and learn the next one without requiring reinitialization.

\section*{\rv{Scalability of learning}}
\rv{Is learning scalable in our metamaterials? To address this question, we systematically explore the learning performance as a function of system size $N$, up to $N=10^3$, in numerical simulations (Supplementary Information Secs.~4-5 and Extended Data Fig.~\ref{figE:convergenceVSsize}). Unsurprisingly, we find that learning becomes harder when $N$ increases. This is due to decay of elastic deformations---this occurs in virtually any elastic structure. Crucially, the metamaterial can still learn by adding longer-range interactions, or by using multiple outputs. To prove this scalability experimentally, we enable the second nearest-neighbor interactions (Eq.~\eqref{eqM:tau_aa}) in a metamaterial consisting of 48 units that can morph into the shape of a cat in response to three inputs (Extended Data Fig.~\ref{figE:cat} and Video S2).}

\rv{Another aspect of scalability is the complexity of the learning rule. Can our metamaterials learn with simpler learning rules? In order to assess how critical the exact form of the learning rule is for convergence, we now test a simplified binarized variant of the contrastive learning rule inspired by \cite{sternSupervisedLearningPhysical2020a} in simulations (see Supplementary Information Sec.~8). Differing from the gradient-based learning rule given by Eq.~\eqref{eq:dkdt}, this alternative only requires measuring whether the angles of output in the clamped state are higher or lower than those in the free state, and the sign of the angles. It is sufficient to train the metamaterial to morph into the same U-shape as in Fig.~\ref{fig:Learning demenstration}c. This highlights that physical learning in our robotic metamaterials does not necessarily require high-precision sensors and complex processors---this is encouraging for future implementations with constrained hardware.}

\rv{In summary,} our metamaterials can learn different shape changes sequentially, \rv{even at large scales and with simpler learning rules}. What would it take to instead learn multiple shapes all at once? In the following, we will show that implementing an extra physical learning rule to evolve non-reciprocal interactions $k_{i}^{a}$ allows our metamaterials to do so.%learn multiple shape changes.

\begin{figure*}[t]
    \centering
    \includegraphics[width=1\textwidth]{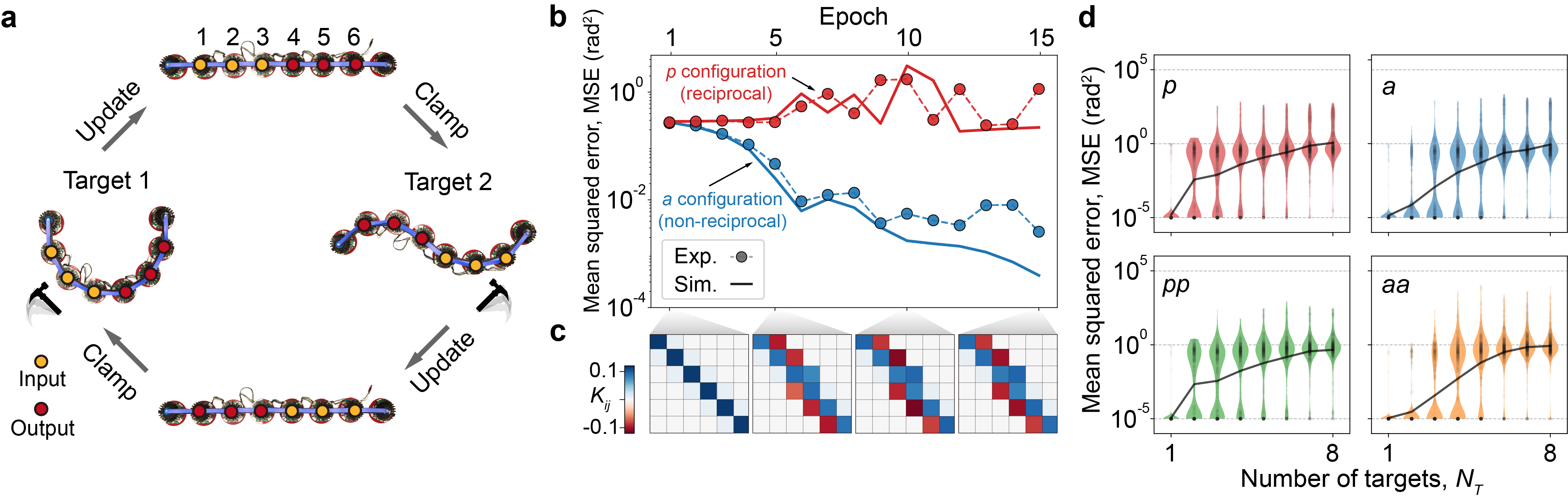}
    \caption{\textbf{Learning non-reciprocal shape changes and multiple targets.} 
    {\bf a,} The procedure of learning non-reciprocal shape changes. Each target shape is learned following the above protocol in Fig.~\ref{fig:Learning demenstration}a but the learning is conducted by switching between these two targets in turn during each epoch. 
    {\bf b,} The MSE curves of learning the above non-reciprocal shape changes in the $p$ configuration (red, \rv{Eq.~\eqref{eq:tau_i} with $k_{i}^{a}=0$}) and the $a$ configuration (blue, \rv{Eq.~\eqref{eq:tau_i} with $k_{i}^{a}\neq0$}) show that these targets can only be learned simultaneously with non-reciprocal interactions, i.e., in the $a$ configuration. Due to human operation error and the precision limitation of the experimental setup, the experimental MSE deviates slightly from the simulated curve after 10 epochs. 
    {\bf c,} The stiffness matrix $K$ of the metamaterial in the  $a$ configuration during learning. The initial parameters are $k_{i}^{o}=0.1$, $k_{i}^{p}=0.01$ and $k_{i}^{a}=0$. The learning rate is $\gamma=0.05$. 
    {\bf d,} Simulation results of learning multiple targets with (non)reciprocal, and next nearest neighbor interactions, i.e., the $pp$ \rv{(Eq.~\eqref{eqM:tau_aa} with $k_{i}^{a}=k_{i}^{aa}=0$)} and $aa$ \rv{(Eq.~\eqref{eqM:tau_aa} with $k_{i}^{a},k_{i}^{aa}\neq0$)} configurations. A system of $N=10$ is simulated and the number of targets $N_{T}$ is varied from 1 to 8. The black semi-transparent dots are the MSE of each simulation and each column consists of 500 simulations. The solid line is the average MSE. The cut-off of the MSE is arbitrarily chosen to be $10^{-5}\ \mathrm{rad}^2$.}
    \label{fig:NR and multi-target}
\end{figure*}

\section*{\rv{Path-dependent contrastive learning rule}}
A non-reciprocal mechanical system eludes the Maxwell-Betti theorem, which stipulates that the transmission of forces into displacements is symmetric with respect to the point of application of the load~\cite{fruchart_odd_2023, veenstra_non-reciprocal_2024, mandal_learning_2024, veenstra_adaptive_2025}. For linear non-reciprocity, the forces do not derive from an energy potential and instead depend on the loading path. If we naively use the elastic energy (Eq.~\eqref{eq:psi_p}) as the function $\psi$, the anti-symmetric terms proportional to $k_{i}^{a}$ are canceled out and do not appear in the learning rule (see Supplementary Information \rv{Sec.~2}). 

To generalize contrastive learning to non-reciprocal systems,  we define a new learning rule that takes into account the path-dependence of the anti-symmetric term $k_{i}^{a}$. To this end, we introduce a path-dependent work instead of the elastic energy as the function $\psi$:
\begin{equation}
\label{eq:work}
\begin{split}
    \psi=&\dfrac{1}{2}\sum_{i=1}^{N}\left(k_{i}^{o}+k^{e}\right)\left(\delta\theta_{i}\right)^{2}\\
     &+\sum_{i=1}^{N-1}\left(k_{i}^{p}\delta\theta_{i}\delta\theta_{i+1}+\alpha_i k_{i}^{a}\delta\theta_{i}\delta\theta_{i+1}\right).
\end{split}
\end{equation}
Here, $\alpha_i=\mathrm{sgn}(i-I)$ for $i\neq I$, or $\alpha_i=\mathrm{sgn}(O-I)$ for $i=I$, which indicates the direction of the loading path between unit $i$ or output unit $O$ and an input unit $I$ (see Supplementary Information \rv{Sec.~2}). \rv{The stiffness $k_{i}^{a}$ to which $\alpha$ applies sets the magnitude of the anti-symmetric torque. This torque changes sign depending on the direction of actuation.} If the $i^\textrm{th}$ unit is on the right side of the input $I$ ($i > I$), the loading path goes from left to right, $\alpha_i=1$ and the contribution to $\psi$ by $k_{i}^{a}$ is positive. In contrast, if the $i^\textrm{th}$ unit is on the left side of the input $I$ ($i < I$), the loading path goes backward from right to left, $\alpha_i=-1$ and the contribution to $\psi$ by $k_{i}^{a}$ is negative. If $i=I$, the contribution of $k_{i}^{a}$ is given by the loading path between output and input units. Substituting Eq.~\eqref{eq:work} into Eq.~\eqref{eq:dkdt}, we obtain the updated values for each stiffness component. The explicit learning rules of $k_{i}^{o}$ and $k_{i}^{p}$ remain the same as Eqs.~\eqref{eq:kodot} and~\eqref{eq:kpdot}, but that of $k_{i}^{a}$ is
\begin{equation}
\label{eq:kadot}
    \frac{\mathrm{d} k_{i}^{a}}{\mathrm{d}t} = -\alpha_i\gamma\left(\delta\theta_{i}^{C}\delta\theta_{i+1}^{C}-\delta\theta_{i}^{F}\delta\theta_{i+1}^{F}\right).
\end{equation}
Now, equipped with this non-reciprocal learning rule by introducing a path-dependent term, we next apply it to our metamaterials to learn non-reciprocal responses. 

\section*{\rv{Learning} non-reciprocal shape changes}
We return to the metamaterial with $N=6$ units and train it to learn the non-reciprocal shape changes depicted in Fig.~\ref{fig:NR and multi-target}a. Specifically, applying a positive curvature to unit 2 leads to a positive curvature to unit 5, whereas applying a positive curvature to unit 5 leads to a negative curvature to unit 2. If one tries to learn this response with a reciprocal metamaterial \rv{(Eq.~\eqref{eq:tau_i} with $k_{i}^{a}=0$, $p$ configuration)}, it fails (Fig.~\ref{fig:NR and multi-target}b), whereas in a nonreciprocal metamaterial \rv{(Eq.~\eqref{eq:tau_i} with $k_{i}^{a}\neq0$, $a$ configuration)}, the learning is successful (Video~S3). This means non-reciprocity is essential for generating shape changes that break the symmetry between loading directions. As learning proceeds, we note that the stiffness matrix of the non-reciprocal metamaterial, which was initially symmetric, gradually becomes asymmetric (Fig.~\ref{fig:NR and multi-target}c). Thus, we can train a reciprocal metamaterial to become non-reciprocal. Such \rv{path-dependent} learning is distinct from all earlier studies on contrastive learning, which only consider reciprocal systems~\cite{stern_supervised_2021, dillavou_demonstration_2022, scellier_energy-based_2023, altman_experimental_2024, dillavou_machine_2024}.

\section*{Multi-target learning}
Non-reciprocity enables the metamaterial to learn multiple shape changes, even if these are not compatible according to the Maxwell-Betti theorem. The question is what sets the maximum number of shape changes?
To answer this question, we systematically learn multiple targets for a $N=10$ metamaterial and compare reciprocal and non-reciprocal cases. We denote the number of targets as $N_{T}$. Here, each target consists of a single randomly selected input unit and a single randomly selected output unit. Similar to Fig.~\ref{fig:NR and multi-target}a, our metamaterial learns these targets in sequence during each epoch to generate all desired shape changes. Our metamaterial performs poorly once the number of targets exceeds one ($N_{T}>1$) in the $p$ configuration (Fig.~\ref{fig:NR and multi-target}d). This is because two distinct shape changes likely break the Maxwell-Betti relation. In contrast, upon introducing $k_{i}^{a}$ (the $a$ configuration) the metamaterial learns well up to $N_{T}=3$. \rv{Furthermore, incorporating the same path-dependent term $\alpha$ into the aforementioned simplified rule, the metamaterial also successfully reproduces the non-reciprocal shape changes in Fig.~\ref{fig:NR and multi-target}a (see Supplementary Information Sec.~8).}

To further increase the number of targets the metamaterial can learn, we consider again scenarios in which the unit cells can also communicate with their next nearest neighbors---we refer to these configurations as $pp$ \rv{(Eq.~\eqref{eqM:tau_aa} with $k_{i}^{a}=k_{i}^{aa}=0$)} and $aa$ \rv{(Eq.~\eqref{eqM:tau_aa} with $k_{i}^{a},k_{i}^{aa}\neq0$)} for the reciprocal and non-reciprocal cases. Whereas the $pp$ configuration does not bring an appreciable improvement, the $aa$ configuration can learn up to $N_{T}=4$. The fact that a larger learning space enables more complex learning tasks is consistent with earlier studies~\cite{rocks_limits_2019, stern_continual_2020} and can be rationalized by a basic constraint counting argument (see Supplementary Information \rv{Sec.~4}). Besides increasing the number of learning degrees of freedom, a straightforward strategy to address this limited learning capacity is to increase the number of units (see Supplementary Information). To illustrate the ability of our metamaterials to learn multiple targets, we train our metamaterial to deform into the letters ``LEREN" (Dutch for ``LEARN") upon application of the appropriate input deformation (Video~S3). In contrast to Fig.~\ref{fig:Learning demenstration}e, there is no retraining, the four letters are learned simultaneously, and the metamaterial can generate all four shapes depending on the angles and locations of input units.

\begin{figure*}[t]
  \centering
  \includegraphics[width=\textwidth]{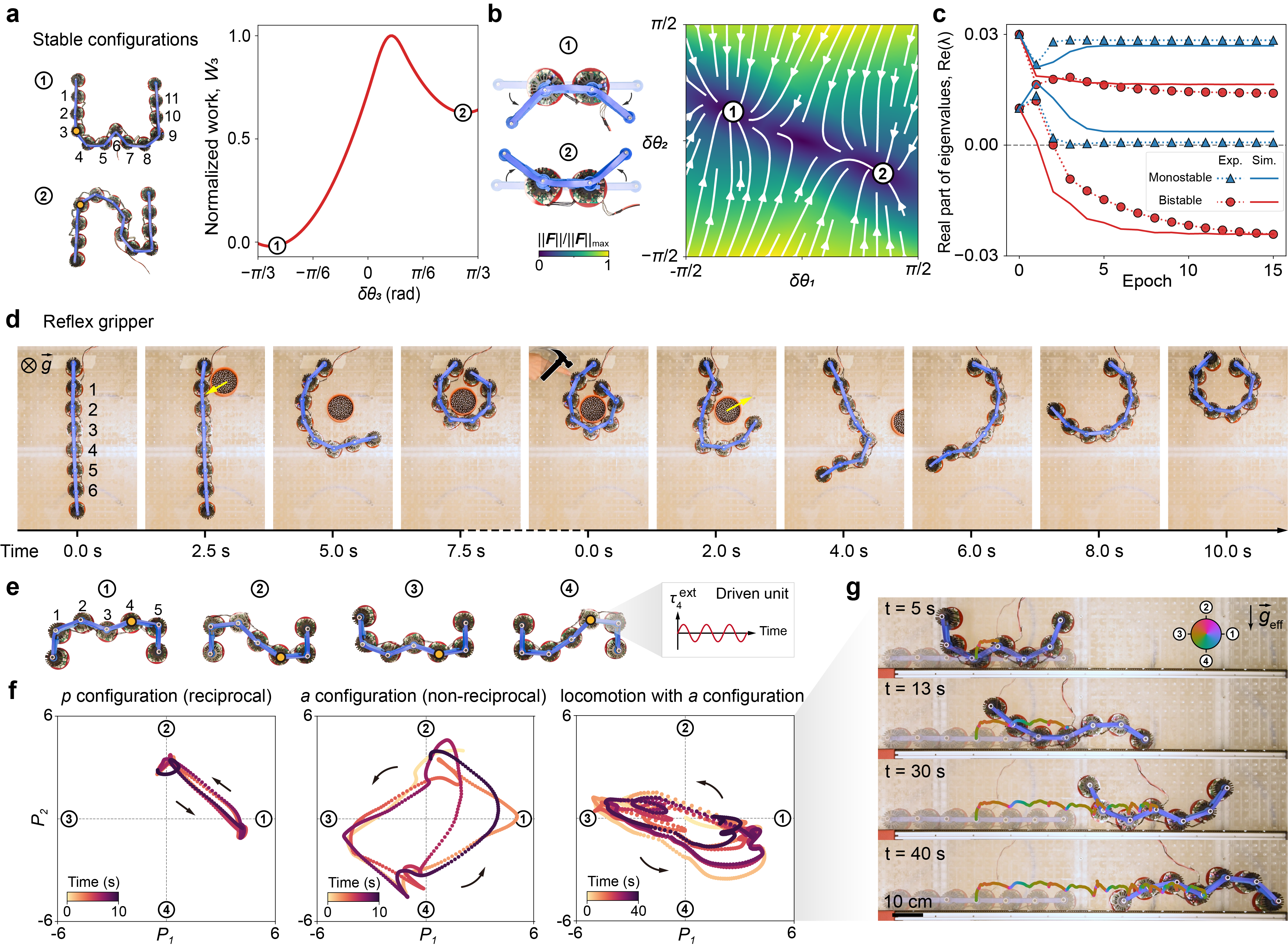}
  \caption{
  \textbf{Learning multistable shape changes and robotic functionalities. } 
    {\bf a,} The normalized work landscape of unit 3 (yellow dot) by tuning $\delta\theta_3$. Two local minima correspond to two stable configurations, the letters ``W" and ``N". 
    {\bf b,} A pair of units shows bistable behavior. The flat configuration corresponds to zero deformations. Upon perturbation, the system jumps to a non-zero deformation instead of springing back to the initial configuration. The left inset shows the two stable configurations. The right inset shows the force fields of the bistable case in which two stable fixed points exist. \rv{The colorbar shows the normalized total torque $||\mathbf{F}||/ ||\mathbf{F}||_{\mathrm{max}}$ where $\mathbf{F}=\{\tau_{1}, \tau_{2}\}^{\top}$ and $\tau_{i}$ is the torque of unit $i$.}
    {\bf c,} The real part of the eigenvalues $\lambda$ during learning for a pair of units in both monostable and bistable scenarios that learn the same target. 
    The desired shape change is generating $\delta\theta_{2}=-\pi/6$ rad after applying $\delta\theta_{1}=\pi/6$ rad. The imaginary part is zero so it is not shown here. 
    {\bf d,} A metamaterial with $N=6$ is trained as a reflex gripper (see Video~S4). It can automatically catch a moving object and release it when an input is applied.
    {\bf e-g,} Using a trained metamaterial with non-reciprocal interactions to achieve locomotion (see Video~S4). 
    {\bf e,} The metamaterial with $N=5$ initially learns to generate the letter ``M" (shape 1) and has four stable shapes. The system is driven by applying an external sine torque $\tau_{4}^{\mathrm{ext}}$ on unit 4 (yellow dot). 
    {\bf f,} The deformation of the system over time with the $p$, $a$ configuration and when it locomotes with the $a$ configuration. Data plotted in shape space projected onto the two \rv{basis} vectors ($P_{1}$, $P_{2}$) (Eq.~\eqref{eqM:cyclic_vectors}) and colored with time. 
    {\bf g,} Snapshots of the locomotion and the trajectory of the center of mass colored by the angle of the projected shape  $P_1+iP_2$. With the airtable inclined under an angle, gravity $\vec{g}_{\text{eff}}$ points downwards.
  }
  \label{fig:Multistable shape-changes}
\end{figure*}

\section*{\rv{Learning} multistable shape changes}
So far, our metamaterials have been trained in monostable scenarios: they spring back to the initial flat configuration once the input units are released. 
Surprisingly, by playing with our metamaterials, we discover that our metamaterials can have multistable configurations (Fig.~\ref{fig:Multistable shape-changes}a and Video~S4). To understand where this unexpected multistability comes from, we start with a pair of units and analyze its stability. Its linear stability is determined by the eigenvalues of the stiffness matrix $K$ \rv{(Extended Data Fig.~\ref{figE:phase_diagram} and see the Supplementary Information Sec.~6)}. The system is unstable if there is at least one negative real eigenvalue. Such negative eigenvalues are made possible by the tunable stiffnesses $k_{i}^{o}$, $k_{i}^{p}$ and $k_{i}^{a}$, which, unlike the stiffness of the elastic skeleton, need not be positive. Therefore, the stiffness matrix need not be positive definite. When one eigenvalue is negative, the deformations amplify exponentially. This amplification is balanced by the limited maximum torque that the motors can apply and the restoring torque from the elastic skeleton. As a result, when the flat configuration is no longer stable, two stable deformed configurations emerge (Fig.~\ref{fig:Multistable shape-changes}b and \rv{Extended Data Fig.~\ref{figE:phase_diagram}b(v)}).

This unexpected discovery triggers a fascinating question: how can we learn multistable shape changes? 
To achieve this, we introduce a local stability constraint to our contrastive learning scheme based on the Gershgorin circle theorem~\cite{semyon_aronovich_uber_1931} (see Supplementary Information \rv{Sec.~7}). \rv{This constraint ensures that local stiffness tuning directly impacts the eigenvalues of the system, which govern its global stability.} In addition, a gradient descent term is added in Eq.~\eqref{eq:kodot}, whose modified version takes the form 
\begin{equation}
\label{eq:GD}
\frac{\mathrm{d}k_{i}^{o}}{\mathrm{d}t}=
-\dfrac{\gamma}{2}\left[\left(\delta\theta_{i}^{C}\right)^{2}-\left(\delta\theta_{i}^{F}\right)^{2}\right]
-2\gamma(k_{i}^{o}-k^{*}).
\end{equation} 
Here, $k^{*}$ is a \rv{local} predetermined value that allows us to tune the \rv{global} stability of the metamaterial. \rv{This extra gradient descent term $2\gamma(k_{i}^{o}-k^{*})$ ensures $k_{i}^{o}$ converges toward $k^{*}$ during training.} For $k^{*}<-k^e$ ($k^{*}>-k^e$), the metamaterial will learn an unstable (stable) shape change provided $|k^*+k^e|>|k_{i-1}^{p}+k_{i-1}^{a}|+|k_{i}^{p}-k_{i}^{a}|$ for any $i$ (for all $i$) (see Supplementary Information \rv{Sec.~7}). Crucially, this constrained learning rule is local and can be implemented with contrastive learning. To prove its feasibility, we use this pair of units and train it to generate the same desired shape changes but with different stability (Fig.~\ref{fig:Multistable shape-changes}c). The eigenvalues always remain positive in the monostable case, while one negative eigenvalue emerges in the bistable case.

Next, we apply this principle to larger metamaterials to achieve robotic functionalities. In Fig.~\ref{fig:Multistable shape-changes}d and Video~S4, we build a reflex gripper that can automatically catch an object once it touches the gripper. Furthermore, the gripper can also release the object and kick it away by pushing unit 1. This is because $k_{1}^{o}$ is negative and unit 1 is bistable. Finally, we use a multistable robotic chain to achieve locomotion. The robotic chain is initially trained to generate the letter ``M''. In order to trigger multistability, $k_{2}^{o}$ and $k_{4}^{o}$ are trained to be negative, so that there are four stable configurations as shown in Fig.~\ref{fig:Multistable shape-changes}e. Surprisingly, the metamaterial exhibits a cyclic shape shift when a sine external torque is applied in a single driven unit (Fig.~\ref{fig:Multistable shape-changes}f and Video~S4)---whereas such cycles are usually achieved with two motors driven with a constant phase delay~\cite{ijspeert_swimming_2007, savoie_robot_2019, oliveri_continuous_2021, li_robotic_2022}. As a result, the metamaterial can locomote on a substrate (Fig.~\ref{fig:Multistable shape-changes}g and Video~S4). We note that such cyclic shape change only occurs when the interactions are non-reciprocal ($a$ configuration, $k_{i}^{a}\neq 0$): \rv{non-reciprocity curls the force field and induces a unidirectional pathway between stable fixed points (Extended Data Fig.~\ref{figE:phase_diagram} b(i, iii))}. Thus, we have shown that periodically driving a single unit generates cycles through shape space by combining multistability~\cite{van_hecke_profusion_2021} and nonreciprocity~\cite{purcell_life_1977, savoie_robot_2019, li_robotic_2022, veenstra_adaptive_2025} which leads to a stable locomotion gait.

\section*{Conclusion}
In conclusion, we have constructed metamaterials that can learn, forget, and relearn to change shape by leveraging a local physical learning strategy. They can do so with multiple shapes, in a nonreciprocal fashion, exhibit multiple stable configurations, and achieve robotic functionalities. Our work paves avenues for the design of adaptive metamaterials~\cite{lee_mechanical_2022, bordiga_automated_2024}, and soft and distributed robotics~\cite{li_particle_2019, oliveri_continuous_2021, saintyves_self-organizing_2024, zou_retrofit_2024, baines_robots_2024}. \rv{Although our current platform focuses on planar shape changes, our learning scheme is inherently general and can be naturally extended to reconfigurable three-dimensional shape-changing metamaterials~\cite{dudek_shape-morphing_2025}. Additional exciting questions ahead are how to extend physical learning to dynamical and stochastic scenarios~\cite{klos_dynamical_2020, mandal_learning_2024, veenstra_adaptive_2025}, and to move from supervised to unsupervised learning. Such advances would further emulate} the autonomous and adaptive behavior of living matter.

\bibliography{refs.bib}% common bib file

% Methodology
\clearpage
\newpage
\beginMethodology
\section*{Methodology}

\begin{figure}[h]
    \centering
    \includegraphics[width=0.6\linewidth]{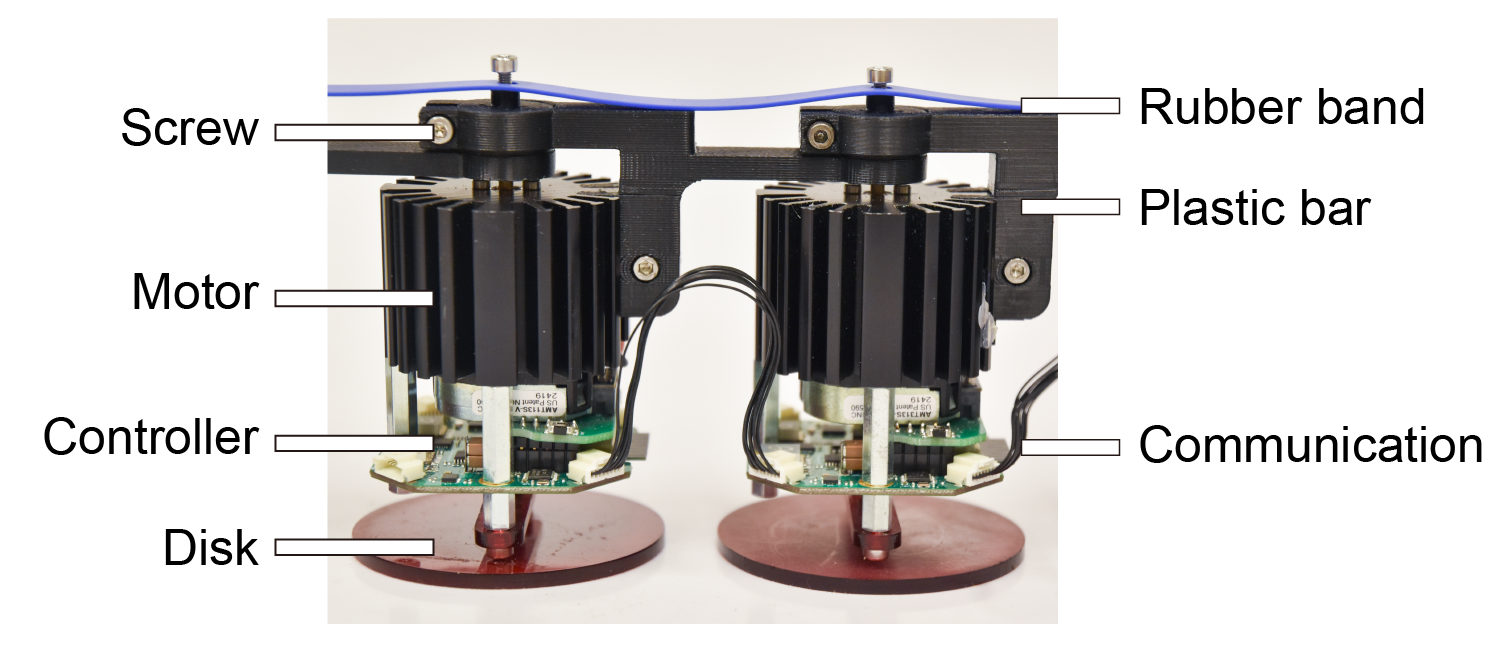}
    \caption{\textbf{The side view of the robotic unit cells.}
    Each unit cell is a motorized vertex connected by 3D printed plastic arms and elastic rubber bands. It consists of a DC motor embedded in a cylindrical heatsink and a microcontroller connected to a custom electronic board. The electronic board enables communication between vertices. Each motorized vertex sits on top of a red circular disk that ensures that the robotic unit floats on the air table. We apply external deformations by manually fastening the screws.}
    \label{figE:robotic unit}
\end{figure}

\vspace{1em}
\noindent\textbf{Experimental protocol}\\
Our robotic metamaterials are made of multiple robotic units composed of motorized vertices connected by 3D printed plastic arms and an elastic skeleton with stiffness $k^{e}$ = 12 $\mathrm{mN}\cdot\mathrm{m/rad}$ (Extended Data Fig.~\ref{figE:robotic unit}). Each vertex consists of a DC coreless motor (Motraxx CL1628) embedded in a cylindrical heatsink, an angular encoder (CUI AMT113S), and a microcontroller (ESP32) connected to a custom electronic board. The electronic board enables power conversion, interfaces the sensor and motor, and enables communication between vertices. Ideally, the motor is programmed to produce an external linear torque as: 
\rv{
\begin{equation}
\label{eqM:tau_linear}
\begin{split}
    \tau_{i}^{\text{linear}} = 
        &-k_{i}^{o}\delta\theta_{i}\\
        &-(k_{i-1}^{p}+k_{i-1}^{a})\delta\theta_{i-1}\\
        &-(k_{i}^{p}-k_{i}^{a})\delta\theta_{i+1}.    
\end{split}
\end{equation}
$k_i^o$, $k_i^p$ and $k_i^a$ are the coupling parameters introduced in the Main Text. $\delta\theta_{i}$ is the angular deflection of $i^{\text{th}}$ unit. In fact, the motor saturates at a maximum torque of $\tau_{\text{max}}$ = 12 $\mathrm{mN}\cdot\mathrm{m}$. So each motor follows a nonlinear force function in practice:
\begin{equation}
\label{eqM:motor torque}
    \tau_{i}^{\text{motor}} = \mathrm{sgn}(\tau_{i}^{\text{linear}})\mathrm{min}(|\tau_{i}^{\text{linear}}|,\tau_{\text{max}}).
\end{equation}
The actual torque of each robotic unit is the sum of the motor torque (Eq.~\eqref{eqM:motor torque}) and the torque generated by the elastic skeleton. It reads:
\begin{equation}
\label{eqM:tau_real}
    \tau_{i} = \tau_{i}^{\text{motor}} - k^{e}\delta\theta_{i}.
\end{equation}}

Experiments are conducted on top of a custom-made, low-friction air table. Each motorized vertex sits on top of a circular disk that ensures that the robotic unit moves on a thin layer of pressurized air smoothly. \rv{However, there is a small amount of residual static friction within the robotic units and between the units and the substrate, even though the setup is designed to minimize such effects. Due to this, there may be slight perturbations in the Video S2, as manual intervention was occasionally applied to help the robot overcome the friction. These interventions are not manual corrections but help the system reach the final configuration. With an ideal, frictionless experimental setup, we expect the metamaterials to morph smoothly and precisely to the trained shape without such interventions. However, this static friction also helps the robotic gripper remain stable in a flat configuration as shown in Fig.~\ref{fig:Multistable shape-changes}d}. The experimental pictures are taken from the top view. By \rv{manually} fastening the screws on the units, we can apply angular deflections on demand. \rv{Though manual clamping is not fundamental to the protocol, and a fully hands-off training could be achieved by using automated actuators to impose target shapes.} The units can store their angular deflections, do calculations in the microcontroller, and update their onsite stiffnesses and neighbor interactions at will.

In Figs.~\ref{fig:Multistable shape-changes}e-g, the airtable was tilted by $1^{\circ}$ with respect to the horizontal plane. This induces an effective gravity $\vec{g}_{\text{eff}}\approx 0.17$~$\mathrm{m\cdot s^{-2}}$ ($\vec{g}\approx 9.78$~$\mathrm{m\cdot s^{-2}}$) pointing toward a treadmill. The frequency of the sinusoidal forcing is 0.25 Hz. The deformation is plotted in the space of two basis deformation vectors, $P_{1}$ and $P_{2}$ defined as
\begin{equation}
\label{eqM:cyclic_vectors}
    P_{1}=\delta\Theta\cdot \frac{\mathbf{v}_{1}}{\|\mathbf{v}_{1}\|}, \
    P_{2}=\delta\Theta\cdot \frac{\mathbf{v}_{2}}{\|\mathbf{v}_{2}\|}.
\end{equation}
Here, $\delta\Theta$ is the angular deflection vector. We define $\mathbf{v}_{1}=\{1, 1,-1,1,1\}^{\top}$ and $\mathbf{v}_{2}=\{1, 1, 0, -1, -1\}^{\top}$ to correspond to the shapes of letter ``M" and letter ``N" respectively in Fig.~\ref{fig:Multistable shape-changes}e.

\vspace{1em}
\noindent\textbf{Simulation protocol}\\
In simulation, we assume the system doesn't saturate so that it always follows a linear force function as Eq.~\eqref{eq:tau_i}. An $N$-unit system follows a constitutive relation as
\begin{equation}
\label{eqM:T}
T = -K\delta\Theta
\end{equation}
where $T=\{\tau_1,\tau_2,\dots,\tau_{N-1},\tau_N\}^\top$ and $\delta\Theta=\{\delta\theta_1,\delta\theta_2,\dots,\delta\theta_{N-1},\delta\theta_N\}^\top$ are the torque and angular deflection vectors of size $N$. $K$ is the stiffness matrix of size $N\times N$. In the case of nearest-neighbor interactions, the above relation is equivalent to:
\begin{widetext}
\begin{equation}
\label{eqM:torque_matrix}
\begin{pmatrix}
\tau_1 \\
\tau_2 \\
\vdots \\
\tau_{N-1} \\
\tau_N
\end{pmatrix}=-
\begin{bmatrix}
k_{1}^{o}+k^{e}     & k_{1}^{p}-k_{1}^{a} & 0                   & 0 & \cdots\\
k_{1}^{p}+k_{1}^{a} & k_{2}^{o}+k^{e}     & k_{2}^{p}-k_{2}^{a} & 0 & \cdots\\
\vdots &   & \ddots                  &                          &  \vdots\\
\cdots & 0 & k_{N-2}^{p}+k_{N-2}^{a} & k_{N-1}^{o}+k^{e}        & k_{N-1}^{p}-k_{N-1}^{a}\\
\cdots & 0 & 0                       & k_{N-1}^{p}+k_{N-1}^{a}  & k_{N}^{o}+k^{e}
\end{bmatrix}
\begin{pmatrix}
\delta\theta_1 \\
\delta\theta_2 \\
\vdots \\
\delta\theta_{N-1}\\
\delta\theta_{N}
\end{pmatrix}.
\end{equation}
\end{widetext}

In contrastive learning~\cite{movellan_contrastive_1991, williamChemicalBoltzmannMachines2017, stern_supervised_2021}, a physical system is trained by observing the contrast between its ``free state" and ``clamped state". For our robotic metamaterials, this procedure follows four steps as shown in Fig.~\ref{fig:Learning demenstration}a, which we now describe in more detail.

(i) \textit{Initialization} --- We set the initial configuration to be flat and ensure that the initial onsite and neighbor interactions are such that the system is monostable (see Supplementary Information). The input and desired output angular deflection vectors are $\delta\Theta^{I}$ and $\delta\Theta^{O}$ of size $N$. The sets of input and output indices are $\mathcal{I}$ and $\mathcal{O}$. 
For example, for a system with $N=3$, if the learning task is to achieve a desired output $\delta\bar\theta_{3}$ once an input $\delta\bar\theta_{1}$ is applied (Fig.~\ref{fig:Learning demenstration}a), then we have $\delta\Theta^{I}=\{\delta\bar\theta_{1}, 0, 0\}^{\top}$, $\delta\Theta^{O}=\{0, 0,\delta\bar\theta_{3}\}^{\top}$, $\mathcal{I}=\{1\}^{\top}$ and $\mathcal{O}=\{3\}^{\top}$. We use $\delta\theta_{i}$ as the $i^{\text{th}}$ entry of the vector $\delta\Theta$ in the following.

(ii) \textit{Free state} --- After applying the input angles $\delta\Theta^{I}$, we calculate the induced torque on each unit $\tau_{i}^{F}$ which is given by
\begin{equation}
\tau_{i}^{F}=-\displaystyle\sum_{j=1}^{N}K_{ij}\delta\theta_{j}^{I}.
\end{equation}
Then, we find the angle vector $\delta\Theta^{F}$ corresponding mechanical equilibrium, i.e., $\tau_{i}=0$ for $i\notin\mathcal{I}$,  by inverting the stiffness matrix $K$. The resulting state is called the free state and reads
\begin{equation}
\delta\theta_{i}^{F}=
\begin{cases}
-\displaystyle\sum_{j=1}^{N}(K^{-1})_{ij}\tau_{j}^{F}, & \text{ if } i \notin \mathcal{I} \\
\delta\theta_{i}^{I},  & \text{ if } i \in \mathcal{I}.
\end{cases}
\end{equation}

(iii) \textit{Clamped state} --- We now determine the nudging angle vector $\delta\Theta^{N}$ with entries
\begin{equation}
\delta\theta_{i}^{N}=
\begin{cases}
\delta\theta_{i}^{F}, & \text{ if } i \notin \mathcal{O} \\
\delta\theta_{i}^{O},  & \text{ if } i \in \mathcal{O},
\end{cases}
\end{equation}
and find the torque on each unit $\tau_{i}^{C}$ induced when the system is clamped at the nudging angle $\delta\Theta^{N}$
\begin{equation}
\tau_{i}^{C}=-\displaystyle\sum_{j=1}^{N}K_{ij}\delta\theta_{j}^{N}.
\end{equation}

We now find the equilibrium configuration of the clamped state given by the angle vector $\delta\Theta^{C}$, whose entries read
\begin{equation}
\delta\theta_{i}^{C}=
\begin{cases}
-\displaystyle\sum_{j=1}^{N}K_{ij}^{-1}\tau_{j}^{C}, & \text{ if } i \notin (\mathcal{I}\cup\mathcal{O}) \\
\delta\theta_{i}^{N},  & \text{ if } i \in (\mathcal{I}\cup\mathcal{O}).
\end{cases}
\end{equation}

(iv) \textit{Updating} --- By substituting the angles of the free $\delta\Theta^{F}$ and clamped states $\delta\Theta^{C}$ into Eqs.~\eqref{eq:kodot}, \eqref{eq:kpdot} and \eqref{eq:kadot}, we update the stiffness matrix $K$. The above operation will be repeated for a number of epochs. The learning error is defined by the mean squared error (MSE):
\begin{equation}
\label{eqM:MSE}
    \text{MSE}=\frac{1}{N_{O}}\sum_{i\in\mathcal{O}}\left(\delta\theta_{i}^{O}-\delta\theta_{i}^{F}\right)^{2},
\end{equation}
where $N_{O}$ is the number of output units. The simulation codes are available in a public Zenodo repository at \cite{du_metamaterials_2025}.

\vspace{1em}
\noindent\textbf{Metamaterials with second nearest-neighbor interactions}\\
In the Main Text, we also consider metamaterials with the next nearest-neighbor interactions. With those interactions, each robotic unit $i$ exerts a torque as follows:
\begin{widetext}
\begin{equation}
\label{eqM:tau_aa}
    \tau_{i}=-\left(k_{i}^{o}+k^{e}\right)\delta\theta_{i}
             -(k_{i-1}^{p}+k_{i-1}^{a})\delta\theta_{i-1}-(k_{i}^{p}-k_{i}^{a})\delta\theta_{i+1}
             -(k_{i-2}^{pp}+k_{i-2}^{aa})\delta\theta_{i-2}-(k_{i-2}^{pp}-k_{i-2}^{aa})\delta\theta_{i+2},
\end{equation}    
\end{widetext}
where $k_{i}^{pp}$ and $k_{i}^{aa}$ are the passive (symmetric) and active (anti-symmetric) next nearest-neighbor stiffnesses. We refer to the case when $k_{i}^{a}=k_{i}^{aa}=0$ as the $pp$ configuration. Otherwise, we refer to the $aa$ configuration.

The path-dependent work $\psi$ for the $aa$ configuration equals
\begin{widetext}
\begin{equation}
\label{eqM:work_aa}
    \psi=\dfrac{1}{2}\sum_{i=1}^{N}\left(k_{i}^{o}+k^{e}\right)\left(\delta\theta_{i}\right)^{2}
         +\sum_{i=1}^{N-1}\left(k_{i}^{p}\delta\theta_{i}\delta\theta_{i+1}+\alpha_i k_{i}^{a}\delta\theta_{i}\delta\theta_{i+1}\right)
         +\sum_{i=1}^{N-2}\left(k_{i}^{pp}\delta\theta_{i}\delta\theta_{i+2}+\alpha_i k_{i}^{aa}\delta\theta_{i}\delta\theta_{i+2}\right).
\end{equation}
\end{widetext}
Substituting Eq.~\eqref{eqM:work_aa} into Eq.~\eqref{eq:dkdt}, the learning rules of $k_{i}^{o}$, $k_{i}^{p}$ and $k_{i}^{a}$ remain the same as Eqs.~\eqref{eq:kodot}, \eqref{eq:kpdot} and \eqref{eq:kadot}, but these of $k_{i}^{pp}$ and $k_{i}^{aa}$ are
\begin{equation}
\label{eqM:kppkaadot}
\begin{split}
    &\frac{\mathrm{d} k_{i}^{pp}}{\mathrm{d}t}=-\gamma\left(\delta\theta_{i}^{C}\delta\theta_{i+2}^{C}-\delta\theta_{i}^{F}\delta\theta_{i+2}^{F}\right), \\
    &\frac{\mathrm{d} k_{i}^{aa}}{\mathrm{d}t}=-\alpha_i\gamma\left(\delta\theta_{i}^{C}\delta\theta_{i+2}^{C}-\delta\theta_{i}^{F}\delta\theta_{i+2}^{F}\right).   
\end{split}
\end{equation}

\vspace{1em}
\noindent\textbf{Data availability}\\
All the data supporting this study are available on the public repository at https://doi.org/10.5281/zenodo.15012427 (ref.~\cite{du_metamaterials_2025}). Source data are provided with this paper.

\vspace{1em}
\noindent\textbf{Code availability}\\
All the codes supporting this study are available on the public repository at https://doi.org/10.5281/zenodo.15012427 (ref.~\cite{du_metamaterials_2025}).

\vspace{1em}
\noindent\textbf{Acknowledgments}\\
We thank M. Stern, V. Vitelli, A. Liu, D. Durian, J. Schwarz, B. Scellier, S. Dillavou, Y. Zhou and J. Binysh for the insightful discussions and suggestions. We thank K. van Nieuwland, D. Giesen, R. Hassing and S. Koot for technical assistance. Y. D. acknowledges financial support from the China Scholarship Council. We acknowledge funding from the European Research Council under Grant Agreement No. 852587 and from the Netherlands Organisation for Scientific Research (NWO) under grant agreement VIDI 2131313.

\vspace{1em}
\noindent\textbf{Author contribution}\\
C. C. and Y. D. conceptualized and guided the project. Y. D. and J. V. designed the samples and experiments. Y. D. carried out the experiments. Y. D. and R. v. M. carried out the numerical simulations. R. v. M. and Y. D. performed the theoretical study. 
All authors contributed extensively to the interpretation of the data and the production of the manuscript. Y. D. and C. C. wrote the main text. Y. D. created the figures and Videos. All authors contributed to the writing of the Methodology and the Supplementary Materials.

\vspace{1em}
\noindent\textbf{Competing interests}\\
There are no competing interests to declare.

\vspace{1em}
\noindent\textbf{Supplementary information}\\
\rv{
Supplementary Sections 1–8, Figures S1–8, Tables S1-2 and Videos S1-4.}

%% Extended Data 
\onecolumngrid
\beginExtendedData
\newpage
\clearpage
\section*{Extended data}

\begin{figure}[h]
    \centering
    \includegraphics[width=0.6\linewidth]{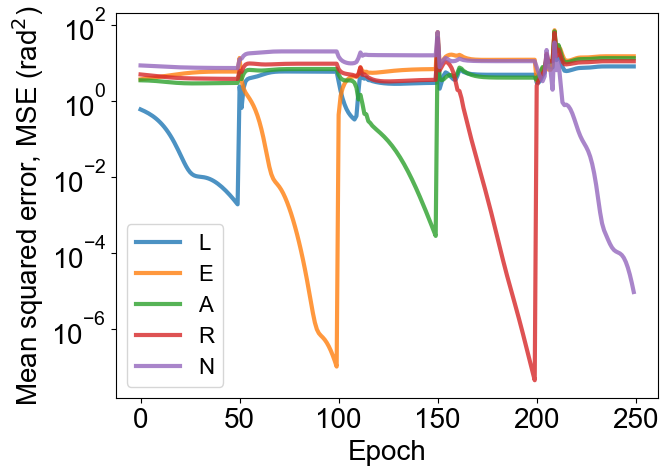}
    \caption{\textbf{The MSE curves of learning to form the word “LEARN” sequentially in Fig.~\ref{fig:Learning demenstration}e.} It shows that metamaterial can forget the previous shape change and relearn the next one without requiring reinitialization. Here, the learning is conducted in simulation and the learning rate is $\gamma=0.01$. The initial parameters are $k_{i}^{o}=0.1$, $k_{i}^{p}=0.01$ and $k_{i}^{a}=0$.}
    \label{figE:continue learning}
\end{figure}

\begin{figure}[h]
    \centering
    \includegraphics[width=1\linewidth]{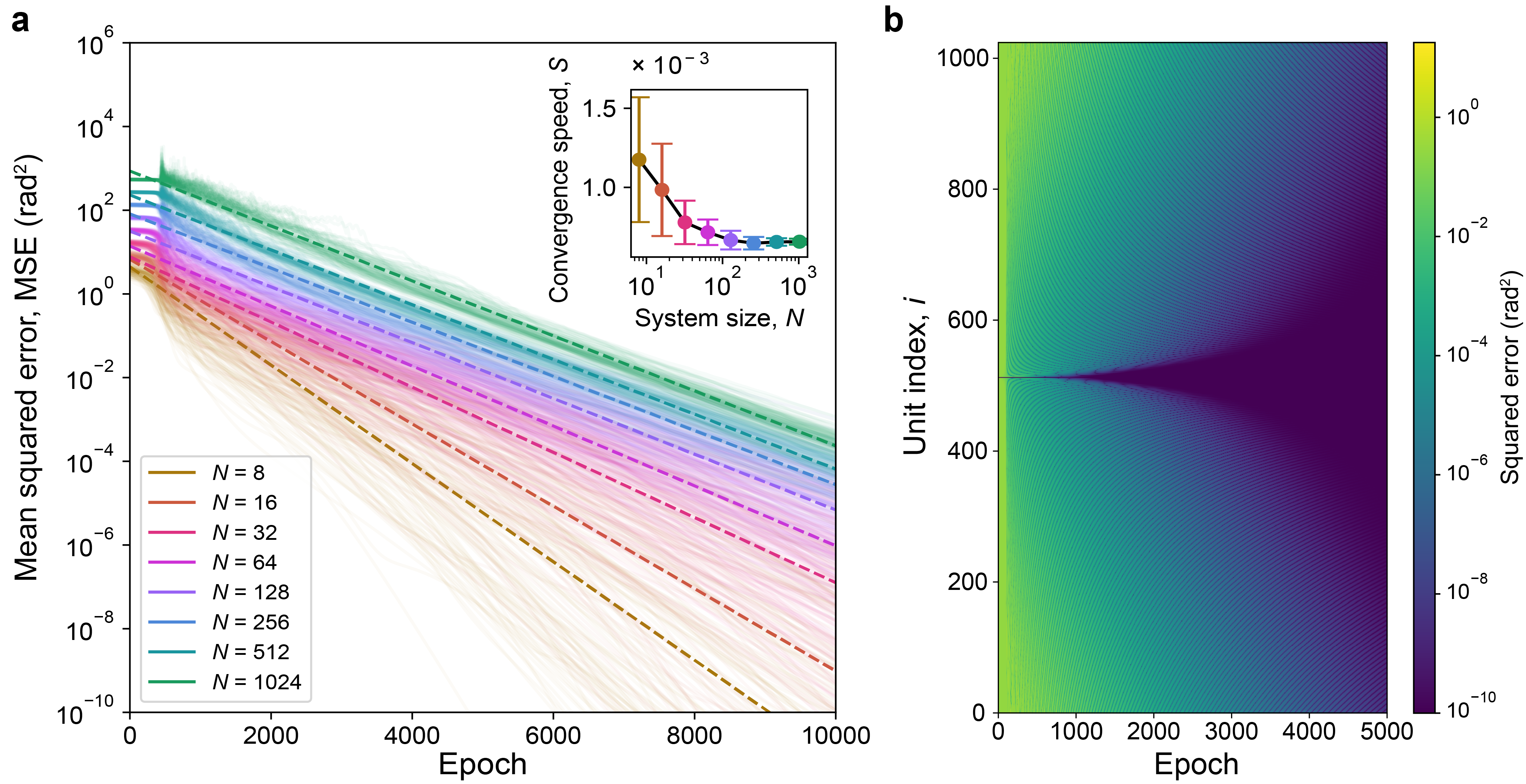}
    \caption{\rv{\textbf{Learning at large scales.}
    \textbf{a,} The MSE curves with varying system size $N$. A system with $N$ units and the second nearest-neighbor interaction ($aa$ configuration, Eq.~\eqref{eqM:tau_aa}) learns a single target that includes one random input and $N-1$ outputs. The solid lines are the MSE curves of each simulation, and there are 100 curves for each system size $N$. The dashed lines are the averaged linear fitting of these 100 error curves. The inset displays the average convergence speed $S$ versus system size $N$. The convergence speed is defined as $S = \frac{1}{N_{s}}\sum_{i}^{N_{s}}S_{i}$ where $N_{s}$ is total number of simulations and $S_{i}$ is the slope of the linear fit for the $i^{\text{th}}$ run. Specifically, $S_{i}$ is extracted from the linear fit of the MSE curve in log scale, $\log(\text{MSE}) = -S_{i}*\text{Epoch}+b_{i}$. $b_{i}$ is the intercept of the fitted line. A higher $S_{i}$ indicates that the learning converges faster. The initial parameters are $k_{i}^{o}=0.1$, $k_{i}^{p}=0.01$, $k_{i}^{a}=0$, $k_{i}^{pp}=0$ and $k_{i}^{aa}=0$. The learning rate $\gamma$ is $10^{-4}$. 
    \textbf{b,} The learning error decays across the system. Here, we consider a system with $N=1024$ units and the $aa$ configuration (Eq.~\eqref{eqM:tau_aa}). It learns a task where the input unit is the $512^{\text{th}}$ unit and the input angle is $\delta\theta_{512}=30^{\circ}$, while all other units are output and their desired angles are $30^{\circ}$. The kymograph shows the squared error $(\delta\theta_{i}^{C}-\delta\theta_{i}^{F})^2$ of each unit during training. It shows that information keeps propagating, which leads to successful learning. Here, the initial parameters are $k_{i}^{o}=0.1$, $k_{i}^{p}=0.01$ and $k_{i}^{a}=0$. The learning rate $\gamma$ is $10^{-3}$. See details in the Supplementary Information Sec.~5.3.1.}}
    \label{figE:convergenceVSsize}
\end{figure}

\begin{figure}[h]
    \centering
    \includegraphics[width=0.8\linewidth]{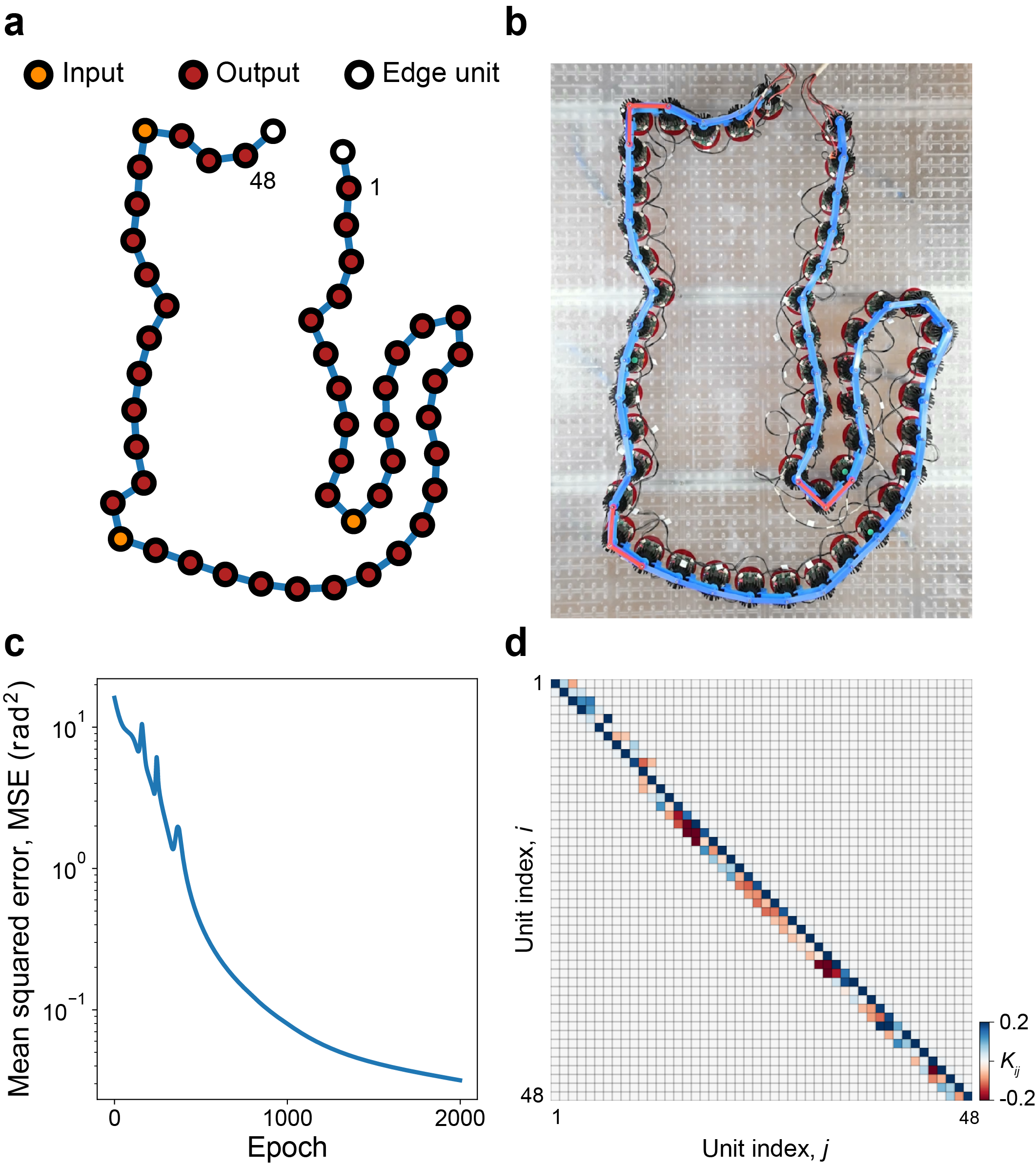}
    \caption{\rv{
    \textbf{Learning to morph into a cat.} A system with $N=48$ units and the second nearest-neighbor interaction ($aa$ configuration, Eq.~\eqref{eqM:tau_aa}) learns to form a shape of cat (see Video~S2). 
    \textbf{a,} The desired shape.
    \textbf{b,} By performing the stiffness matrix obtained in simulation (d), the final learned system morphs into a cat in response to 3 inputs in the experiment. The red linkage applies the input angular deflection.
    \textbf{c,} The corresponding MSE curves during learning.
    \textbf{d,} The stiffness matrix $K$ after learning. Here, the initial parameters are $k_{i}^{o}=0.1$, $k_{i}^{p}=0.01$, $k_{i}^{a}=0$, $k_{i}^{pp}=0$ and $k_{i}^{aa}=0$. The learning rate $\gamma$ is $10^{-4}$.}
    }
    \label{figE:cat}
\end{figure}

\begin{figure}[h]
    \centering
    \includegraphics[width=1\linewidth]{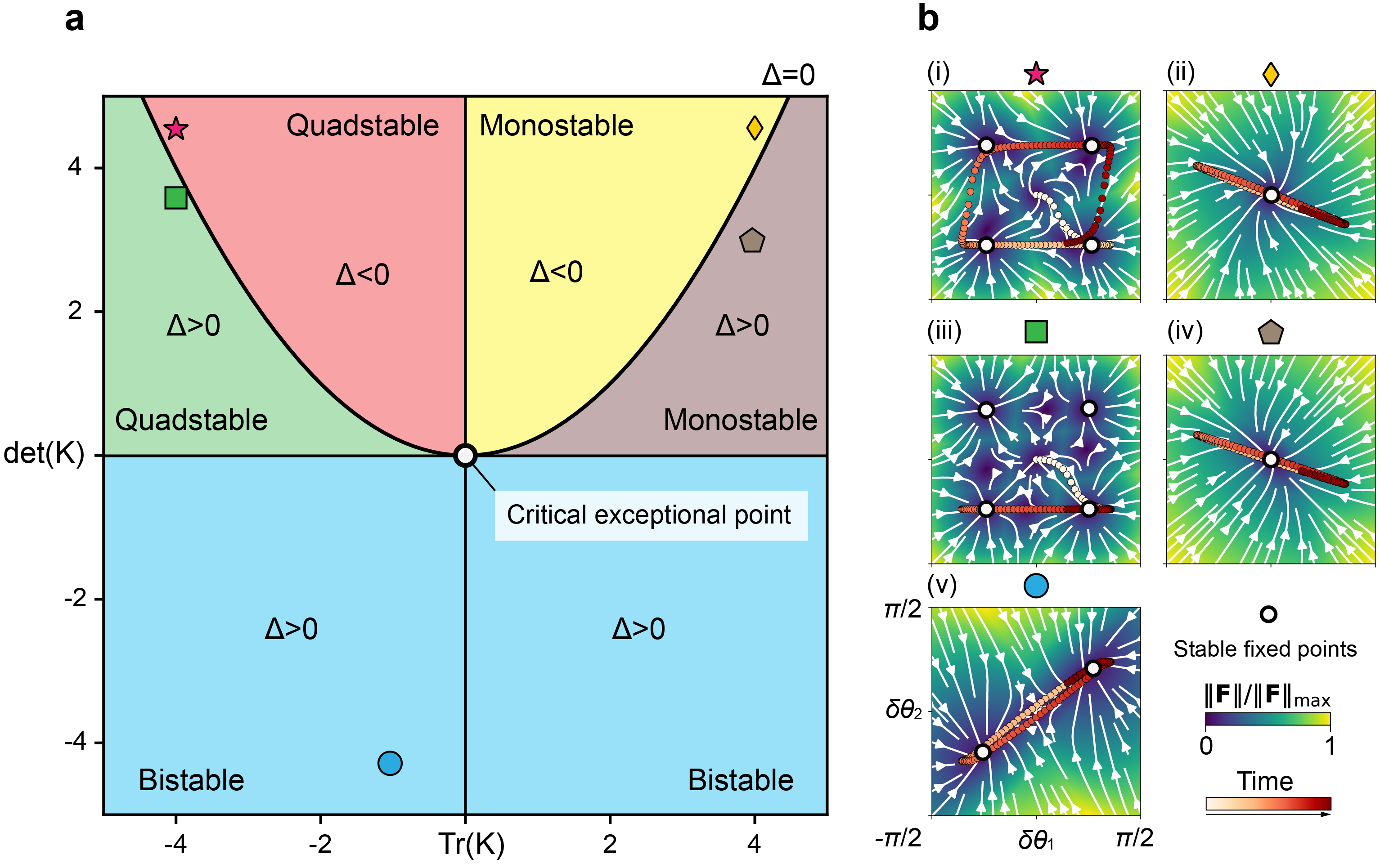}
    \caption{\rv{
    \textbf{Overdamped stability of a 2-unit system.} Because the metamaterials learn static shape changes, we consider a 2-unit system with overdamped dynamics, i.e., 
    $\binom{\delta\dot{{\theta}}_{1}}{\delta\dot{\theta}_{2}}=
    -\begin{bmatrix}
     k_{1}^{o}   & k_{1}^{p}-k_{1}^{a} \\ 
     k_{1}^{p}+k_{1}^{a} & k_{2}^{o}
    \end{bmatrix}
    \binom{\delta\theta_{1}}{\delta\theta_{2}}$, and analyze the stability while considering nonlinear motor saturation (Eq.~\eqref{eqM:motor torque}).
    \textbf{a,} The stability phase diagram of this 2-unit system. It is plotted in the space of the determinant $\mathrm{det}(K)$ and trace $\mathrm{Tr}(K)$ of the stiffness matrix. $\Delta=\mathrm{Tr}(K)^2-4\mathrm{det}(K)$. The eigenvalue $\lambda$ of the stiffness matrix determines the linear stability of this 2-unit system and $\lambda=\frac{1}{2}[\mathrm{Tr}(K)\pm\sqrt{\Delta}]$. In the red, yellow, green, brown and blue regions, $\lambda$ are two complex numbers with a negative real part, two complex numbers with a positive real part, two real negative numbers, two real positive numbers, one positive and one negative real number, respectively. These regimes encircle a critical exceptional point (white dot)~\cite{al-izziNonreciprocalBucklingMakes2025}. It separates the monostable and multistable phases, at which point the eigenvalues and eigenvectors coalesce. (See details in the Supplementary Information Sec.~6).
    \textbf{b,} Force fields corresponding to example points with nonlinear motor saturation (Eq.~\eqref{eqM:motor torque}). The color map indicates the torque magnitude normalized by the maximum torque. The overlaid dots show the trajectory of the corresponding overdamped system under a sine torque applied to unit 1.}}
    \label{figE:phase_diagram}
\end{figure}

\clearpage
\newpage
\beginSI
\section*{Supplementary Information}
\vspace{10pt}
\section{List of supplementary videos}
Description of supplementary videos:

\begin{itemize}
\item Supplementary Video S1: \textbf{Summary}.
In this video, we summarize building a robotic metamaterial that learns to shape change by using a contrastive learning scheme. Our metamaterial can learn shape changes sequentially, non-reciprocal responses, \rv{where multiple shapes are incompatible in equilibrium}, and even multistable shape changes. All these capabilities taken together enable robotic functionalities.

\item Supplementary Video S2: \textbf{Learning procedure and complex learned shape changes}.
We introduce details of our contrastive learning procedure by giving an example of learning to form the letter ``U". We show our metamaterial can sequentially learn complex shape changes by training a metamaterial with 11 units to form the word ``LEARN". \rv{Eventually, we show a metamaterial with 48 units morphs into the shape of a cat.}

\item Supplementary Video S3: \textbf{Learning non-reciprocal and multiple shape changes}.
We demonstrate that our learning scheme can successfully learn non-reciprocal shape changes. Furthermore, by including asymmetric and next nearest neighbor interactions, our metamaterial learns multiple shape changes simultaneously. Concretely, we train a metamaterial with 11 units to learn to form the word ``LEREN" (Dutch for ``LEARN”) upon application of the appropriate input deformations.

\item Supplementary Video S4: \rv{\textbf{From multistable shape changes to functionality.}}
We show the experimental discovery of multistability in our metamaterials and two functional examples by learning multistable shape changes. In the first example, we train a bistable metamaterial to perform reflex gripping actions. This gripper can both automatically catch an object once it touches the gripper, but also release and kick it away by pushing one unit. In the second example, we train a metamaterial to be multistable and it exhibits a cyclic shape shift when introducing non-reciprocity and a sine external torque is applied in a single driven unit. This allows the metamaterial to locomote on a substrate.
\end{itemize}

\section{Derivation of path-dependent contrastive learning rule}
In an earlier study~\cite{stern_supervised_2021}, contrastive learning was applied to passive, reciprocal systems, using a learning rule derived from the elastic energy difference between the free and clamped states. If we consider a system described by Eq.~(1), its elastic energy $E$ takes the following form:
\begin{equation}
    E=-\frac{1}{2}\delta\Theta^{\top}K\delta\Theta\\
    =-\frac{1}{2}\displaystyle\sum_{i=1}^{N}(k_{i}^{o}+k^{e})(\delta\theta_{i})^{2}-\displaystyle\sum_{i=1}^{N-1}k_{i}^{p}\delta\theta_{i}\delta\theta_{i+1},
\end{equation}
where $\delta\Theta$ is the angular deflection vector and $K$ is the stiffness matrix. Crucially, note that the active stiffness $k_{i}^{a}$ does not contribute to the total elastic energy. Thus, the elastic energy alone is insufficient to devise an update rule for $k_{i}^{a}$.

To generalize contrastive learning to non-reciprocal systems, we use a new function $\psi$ as shown in Eqs.~(6) and \eqref{eqS:psi} based on path-dependent work. First, we derive the path-dependent work in a 2-unit system, then we generalize it to an $N$-unit system and finally generalize it to our new contrastive learning rule. Note that the motor saturation (Eq.~(M2)) is not considered here.

\subsection{Path-dependent work of a 2-unit system}
We now consider a 2-unit system and its constitutive relation is
\begin{equation}
\label{eqS:2unitsystem}
    \binom{\tau_{1}}{\tau_{2}}=-
    \begin{bmatrix}
     k_{1}^{o} + k^{e}   & k_{1}^{p}-k_{1}^{a}\\ 
     k_{1}^{p}+k_{1}^{a} & k_{2}^{o}+k^{e}
    \end{bmatrix}\binom{\delta\theta_{1}}{\delta\theta_{2}}.
\end{equation}
We intend to train unit $2$ to deform in response to an input deflection of unit $1$ as $\delta\bar\theta_{1}\rightarrow \delta\bar\theta_{2}$. To learn this response, we first apply $\delta\bar\theta_{1}$, and allow the system to reach the corresponding free state, given by mechanical equilibrium. The work done to reach the free state is called $W_{1\rightarrow 2}^{F}$.
We then clamp the system by nudging $\delta\theta_2$ to its desired response $\delta\bar\theta_{2}$ while keeping $\delta\theta_1$ fixed as $\delta\bar\theta_1$. The work done by nudging the system to the clamped state from the free state is $\Delta W_{1\rightarrow2}$ and the work to achieve the clamped state from the initial configuration is $W_{1\rightarrow2}^{C}$.
We assume the loading is applied quasi-statically and that the instantaneous torque is the only force that does work when the system equilibrates to the free or clamped state. Explicitly, the work terms are
\begin{equation}
\label{eqS:DWF12}
W_{1\rightarrow2}^{F}=\int_{0}^{\delta{\theta}_{1}^{F}}\tau_{1}\mathrm{d}\delta\theta_{1}+\int_{0}^{\delta{\theta}_{2}^{F}}\tau_{2}\mathrm{d}\delta\theta_{2},
\end{equation}

\begin{equation}
\label{eqS:DWC12}
W_{1\rightarrow2}^{C}=\int_{0}^{\delta{\theta}_{1}^{C}}\tau_{1}\mathrm{d}\delta\theta_{1}+\int_{0}^{\delta{\theta}_{2}^{C}}\tau_{2}\mathrm{d}\delta\theta_{2},
\end{equation}

\begin{equation}
\label{eqS:DW12}
\Delta W_{1\rightarrow2}=\int_{\delta{\theta}_{1}^{F}}^{\delta{\theta}_{1}^{C}} \tau_{1}\mathrm{d}\delta\theta_{1}+\int_{\delta{\theta}_{2}^{F}}^{\delta{\theta}_{2}^{C}} \tau_{2}\mathrm{d}\delta\theta_{2}.
\end{equation}

$W_{1\rightarrow2}^{F}$ is straightforward to evaluate since $\tau_{2}=0$ and thus only $\tau_{1}$ does work in the free state. However,$W_{1\rightarrow2}^{C}$ cannot be evaluated directly because both $\tau_{1}$ and $\tau_{2}$ do work and are functions of $\delta\theta_{1}$ and $\delta\theta_{2}$: this work requires an explicit loading path to calculate the integral Eq.~\eqref{eqS:DWC12}. Fortunately, the work difference $\Delta W_{1\rightarrow2}$ is again straightforward to evaluate because $\delta{\theta}_{1}^{F}=\delta{\theta}_{1}^{C}=\delta\bar\theta_{1}$, such that Eq.~\eqref{eqS:DW12} simplifies to
\begin{equation}
\label{eqS:DW12_extended}
\begin{split}
    \Delta W_{1\rightarrow2}=
    &\int_{\delta{\theta}_{2}^{F}}^{\delta{\theta}_{2}^{C}} \tau_{2}\mathrm{d}\delta\theta_{2}\\
    =&\int_{\delta{\theta}_{2}^{F}}^{\delta{\theta}_{2}^{C}}\left[-(k_{1}^{p}+k_{1}^{a})\delta\theta_{1}^{F}-(k_{2}^{o}+k^{e})\delta\theta_{2}\right]\mathrm{d}\delta\theta_{2}\\
    =&-\frac{1}{2}(k_{2}^{o}+k^{e})\left[(\delta\theta_{2}^{C})^{2}-(\delta\theta_{2}^{F})^{2}\right]
    -(k_{1}^{p}+k_{1}^{a})(\delta\theta_{1}^{C}\delta\theta_{2}^{C}-\delta\theta_{1}^{F}\delta\theta_{2}^{F}).
\end{split}
\end{equation}

Conversely, if we intend to learn the reciprocal target $\delta\bar\theta_{2}\rightarrow \delta\bar\theta_{1}$, the work difference $\Delta W_{2\rightarrow1}$ equals
\begin{equation}
\label{eqS:DW21}
    \Delta W_{2\rightarrow1}=\int_{\delta{\theta}_{1}^{F}}^{\delta{\theta}_{1}^{C}} \tau_{1}\mathrm{d}\delta\theta_{1}+\int_{\delta{\theta}_{2}^{F}}^{\delta{\theta}_{2}^{C}} \tau_{2}\mathrm{d}\delta\theta_{2}.
\end{equation}
Using $\delta\theta_{2}^{F}=\delta\theta_{2}^{C}=\delta\bar\theta_{2}$, Eq.~\eqref{eqS:DW21} simplifies to
\begin{equation}
\label{eqS:DW21_extended}
\begin{split}
    \Delta W_{2\rightarrow1}=&\int_{\delta{\theta}_{1}^{F}}^{\delta{\theta}_{1}^{C}} \tau_{1}\mathrm{d}\delta\theta_{1}\\
    =&\int_{\delta{\theta}_{1}^{F}}^{\delta{\theta}_{1}^{C}}\left[-(k_{1}^{o}+k^{e})\delta\theta_{1}-(k_{1}^{p}-k_{1}^{a})\delta\theta_{2}^{F}\right]\mathrm{d}\delta\theta_{1}\\
    =&-\frac{1}{2}(k_{1}^{o}+k^{e})\left[(\delta\theta_{1}^{C})^{2}-(\delta\theta_{1}^{F})^{2}\right]
     -(k_{1}^{p}-k_{1}^{a})(\delta\theta_{1}^{C}\delta\theta_{2}^{C}-\delta\theta_{1}^{F}\delta\theta_{2}^{F}).
\end{split}
\end{equation}

\begin{figure}[htbp!]
    \centering
    \includegraphics[width=1\linewidth]{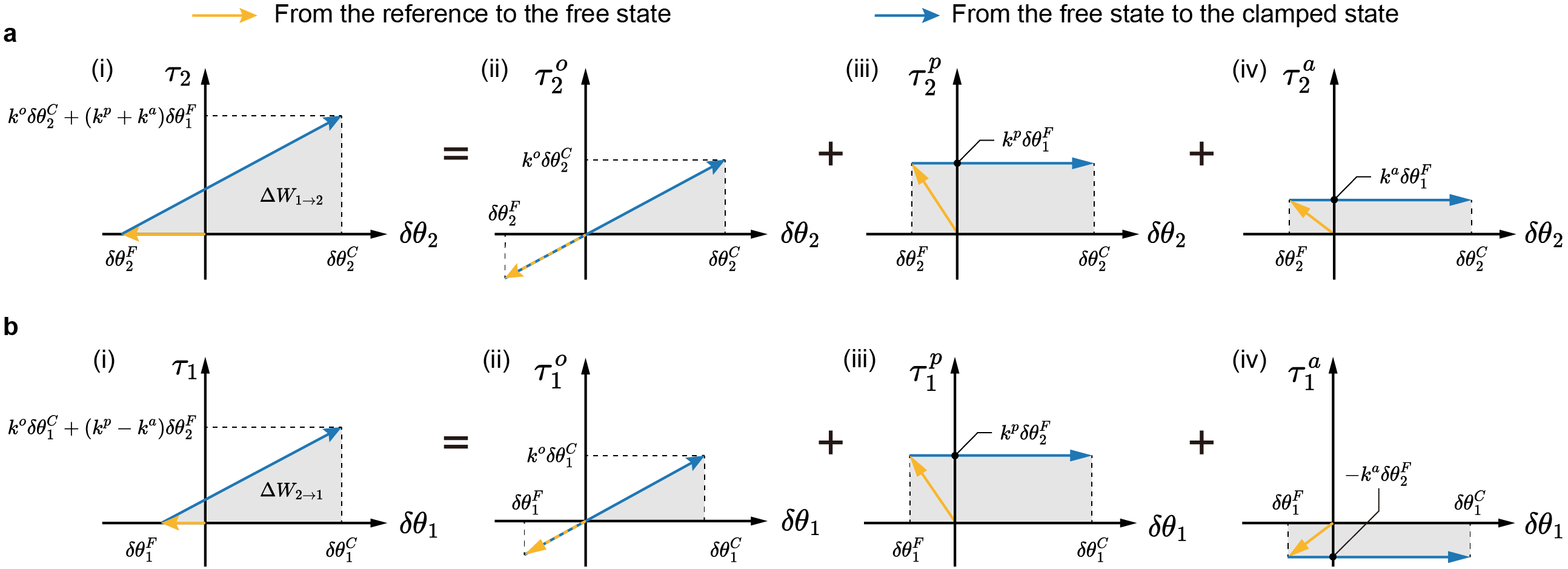}
    \caption{
    \rv{
    \textbf{Torque-angular deflection diagram of a 2-unit system.} We consider a 2-unit system with $k_{i}^{o}=k^{o}$, $k_{i}^{p}=k^{p}$, $k_{i}^{a}=k^{a}$ and $k^{e}=0$ for simplicity. (a-b) shows the two loading paths:  $1\rightarrow 2$ and $2 \rightarrow 1$, respectively. We assume quasi-static loading. $\tau^{o}$, $\tau^{p}$ and $\tau^{a}$ indicate the torque related to onsite, symmetric and anti-symmetric stiffness components ($k^o$, $k^p$ and $k^a$). The yellow and blue arrows show the loading directions of the free state and clamped state. The gray shaded area indicates the work difference $\Delta W$ between the free state and clamped state. We split the total work (i) into contributions from the onsite (ii), symmetric (iii) and anti-symmetric (iv) torques.
    \textbf{a, } During the $1\rightarrow 2$ loading path, only $\tau_{2}$ does work. In the free state, unit 1 is clamped to $\delta\theta_{1}^{F}$ and unit 2 rotates to $\delta\theta_{2}^{F}$. In the clamped state, unit 1 is fixed to $\delta\theta_{1}^{F}$ and unit 2 is clamped to $\delta\theta_{2}^{C}$. Here, only $\tau_{2}$ does work. We note that the work done by $k^a$ is positive here. 
    \textbf{b, } During the $2\rightarrow 1$ loading path, only $\tau_{1}$ does work. In the free state, unit 2 is clamped to $\delta\theta_{2}^{F}$ and unit 1 rotates to $\delta\theta_{1}^{F}$. In the clamped state, unit 2 is fixed to $\delta\theta_{2}^{F}$ and unit 1 is clamped to $\delta\theta_{2}^{C}$. Here, only $\tau_{1}$ does work. Note that the work done by $k^a$ is negative here.}
    }
    \label{figS:path_dependent_work}
\end{figure}

Comparing Eqs.~\eqref{eqS:DW12_extended} and \eqref{eqS:DW21_extended}, we can see the contribution of $k_{i}^{a}$ is path dependent. We combine Eqs.~\eqref{eqS:DW12_extended} and \eqref{eqS:DW21_extended} and now define $\Delta W$ as the path-dependent work difference between the free state and the clamped state. In this case, $\Delta W$ equals
\begin{equation}
\label{eqS:DW_2unit}
    \Delta W=-\frac{1}{2}(k_{1}^{o}+k^{e})\left[(\delta\theta_{1}^{C})^{2}-(\delta\theta_{1}^{F})^{2}\right]
             -(k_{1}^{p}+\alpha k_{1}^{a})(\delta\theta_{1}^{C}\delta\theta_{2}^{C}-\delta\theta_{1}^{F}\delta\theta_{2}^{F}).
\end{equation}
Here, $\alpha=\pm1$ indicates the direction of the loading path. \rv{The $\alpha$ parameter can be understood intuitively as follows. The stiffness $k_{i}^{a}$ to which $\alpha$ applies sets the magnitude of the anti-symmetric torque. This torque changes sign depending on the direction of actuation, viz., the anti-symmetric torque on unit 1 due to a rotation in unit 2 is equal in magnitude and opposite in sign to the anti-symmetric torque on unit 2 due to a rotation in unit 1. Thus, when we consider the work done by moving the output node from the free to the clamped state, the contribution of the anti-symmetric torque to this work will change sign depending on whether the input-output relation is $1\rightarrow 2$ or $2\rightarrow 1$. In detail, for the learning targets $\delta\bar\theta_{1}\rightarrow \delta\bar\theta_{2}$ and $\delta\bar\theta_{2}\rightarrow \delta\bar\theta_{1}$, the loading paths are $\text{unit 1} \rightarrow \text{unit 2}$ and $\text{unit 2} \rightarrow \text{unit 1}$, and $\alpha=1 \text{ and } -1$ respectively. Concretely, this can be seen when we plot the work for these two relations in Fig.~\ref{figS:path_dependent_work}. The work (Eq.~\eqref{eqS:DW_2unit}) is thus path-dependent and we note that it is consistent with $\psi^{C}-\psi^{F}$.}

\subsection{Path-dependent work of an $N$-unit system}
To explain the rationale behind the path-dependent work in the general case, we now derive the path-dependent work in the $N$-unit system with a constitutive relation Eq.~(M6). We first consider the case with a single input and output, then we generalize it to the case with multiple inputs and outputs.

\subsubsection{Single input and output}
We train an output unit $O$ to deform in response to an input deflection of unit $I$: $\delta\bar\theta_{I}\rightarrow \delta\bar\theta_{O}$. To this end, we apply $\delta\bar\theta_{I}$ and allow the system to reach the corresponding free state at mechanical equilibrium. We then clamp the system by nudging $\delta\theta_{O}$ to its desired response $\delta\bar\theta_{O}$ while keeping $\delta\theta_{I}$ fixed to $\delta\bar\theta_{I}$. We first consider the case that the input unit is on the left of the output unit, i.e., $I<O$. The work done by nudging the system to the clamped state from the free state is referred to as $\Delta W$. We assume the nudging is quasi-static and thus $\Delta W$ is equal to
\begin{equation}
\label{eqS:DW_integral}
    \Delta W  
     = \sum_{i}^{N}\int_{\delta{\theta}_{i}^{F}}^{\delta{\theta}_{i}^{C}} \tau_{i}\mathrm{d}\delta\theta_{i}
     = \int_{\delta{\theta}_{O}^{F}}^{\delta{\theta}_{O}^{C}} \tau_{O}\mathrm{d}\delta\theta_{O}
     =\int_{\delta{\theta}_{O}^{F}}^{\delta{\theta}_{O}^{C}}\left(-k^{+}_{O-1}\delta\theta_{O-1}-k^{o}_{O}\delta\theta_{O}-k_{O}^{-}\delta\theta_{O+1}\right)\mathrm{d}\delta\theta_{O}.
\end{equation}
Here, we denote $k_{i}^{p}\pm k_{i}^{a}$ as $k_{i}^{\pm}$ and set $k^e=0$ for convenience, yet without loss of generality. At mechanical equilibrium, all torques are zero except $\tau_I$ and $\tau_{O}$, since this is where external torques are applied to the system. In addition, $\delta{\theta}_{I}^{F}=\delta{\theta}_{I}^{C}$ since the same deformation is applied to unit $I$ in the free state and clamped state. Therefore $\int\tau_{O}\mathrm{d}\delta\theta_{O}$ is the only term left in Eq.~\eqref{eqS:DW_integral}. We note that $\delta\theta_{O-1}$ and $\delta\theta_{O+1}$ depend on $\delta\theta_{O}$. In order to calculate this integral, we need to derive expressions $\delta\theta_{O-1}(\delta\bar\theta_{I}, \delta\theta_{O})$ and $\delta\theta_{O+1}(\delta\theta_{O})$. 

For this, we use two reduced stiffness matrices $K^{L}=K_{I+1:O-1,I+1:O-1}$ and $K^{R}=K_{O+1:N,O+1:N}$. The superscript $L$ ($R$) refers to the left (right) side of $O$. We use index slicing notation where $A^*=A_{i:j}$ denotes that we take $A^*$ to be equivalent to the $A$ matrix taken from index $i$ to index $j$. We first find the expression $\delta\theta_{O-1}(\delta\bar\theta_{I}, \delta\theta_{O})$. We find $\delta\Theta^{L}$=$\delta\Theta_{I+1:O-1}$ by solving
\begin{equation}
    \delta\Theta^{L}=-(K^{L})^{-1}T^{L}.
\end{equation}
Here, $T^{L}$ and $\delta\Theta^{L}$ refer to the reduced torque and angular deflection vector of size $O-I-1$. The entries of $T^{L}$ read
\begin{equation}
\label{eqS:tau_L}
    \tau_{i}^{L}=
    \begin{cases}
     k_{I}^{+}\delta\bar\theta_{I} & \text{ if } i=I+1 \\
     k_{O-1}^{-}\delta\theta_{O} & \text{ if } i=O-1\\
     0 & \text{else.}
    \end{cases}
\end{equation}
Thus, we obtain
\begin{equation}
\label{eqS:dt_O-1}
\begin{split}
    \delta\theta_{O-1}^{L} 
    = &-\sum_{i=I+1}^{O-1}(K^{L})_{O-1,i}^{-1}\tau_{i}^{L}\\
    = &-(K^{L})_{O-1,I+1}^{-1}\tau_{I+1}^{L}-(K^{L})_{O-1,O-1}^{-1}\tau_{O-1}^{L}\\
    = &-(K^{L})_{O-1,I+1}^{-1}k_{I}^{+}\delta\bar\theta_{I}-(K^{L})_{O-1,O-1}^{-1}k_{O-1}^{-}\delta\theta_{O}.
\end{split}
\end{equation}

Next, we find the expression $\delta\theta_{O+1}(\delta\theta_{O})$. Similarly, we find $\delta\Theta^{R}$=$\delta\Theta_{O-1:N}$ by solving
\begin{equation}
    \delta\Theta^{R}=-(K^{R})^{-1}T^{R}.
\end{equation}
Here, $T^{R}$ and $\delta\Theta^{R}$ refer to the reduced torque and angular deflection vector of size $N-O$. The entries of $T^{R}$ read as
\begin{equation}
\label{eqS:tau_R}
    \tau_{i}^{R}=
    \begin{cases}
     k_{O}^{+}\delta\theta_{O} & \text{ if } i=O+1 \\
     0 & \text{ else.}
    \end{cases}
\end{equation}
Thus, we obtain
\begin{equation}
\label{eqS:dt_O+1}
    \delta\theta_{O+1}^{R}=-\sum_{i=O+1}^{N}(K^{R})_{O+1,i}^{-1}\tau_{i}^{R}=-(K^{R})_{O+1,O+1}^{-1}\tau_{O+1}^{R}=-(K^{R})_{O+1,O+1}^{-1}k_{O}^{+}\delta\theta_{O}.
\end{equation}

Substituting Eqs.~\eqref{eqS:dt_O-1} and~\eqref{eqS:dt_O+1} into Eq.~\eqref{eqS:DW_integral}, we have 
\begin{equation}
\label{eqS:DW_withK}
\begin{split}
    \Delta W = \big[ 
    & -\frac{1}{2}k_{O}^{o}(\delta\theta_{O})^{2} \\
    & +k_{O-1}^{+}(K^L)^{-1}_{O-1, I+1}k_{I}^{+}\delta\bar\theta_{I}\delta\theta_O
      +\frac{1}{2}k_{O-1}^{+}(K^L)^{-1}_{O-1,O-1}k_{O-1}^{-}(\delta\theta_{O})^{2}\\
    & +\frac{1}{2}k_{O}^{-}(K^R)^{-1}_{O+1,O+1}k_{O}^{+}(\delta\theta_{O})^{2} \big]
    \bigg|_{\delta\theta_O^F}^{\delta\theta_O^C}.
\end{split}
\end{equation}

The above equation can be simplified to
\begin{equation}
\label{eqS:DW_explicit}
\begin{split}
    \Delta W = 
    & -\frac{1}{2}k_{O}^{o}[(\delta\theta_{O}^{C})^2 - (\delta\theta_{O}^{F})^2] 
    - \frac{1}{2}\left(\prod_{i=I+1}^{O-1}\frac{k_{i}^{+}}{k_{i}^{-}}\right)k_{I}^{+} (\delta\theta_{I}^{C}\delta\theta_{I+1}^{C} -\delta\theta_{I}^{F}\delta\theta_{I+1}^{F})\\
    & -\frac{1}{2}k_{O-1}^{+}(\delta\theta_{O-1}^{C}\delta\theta_{O}^{C} -\delta\theta_{O-1}^{F} \theta_{O}^{F})
      -\frac{1}{2}k_{O}^{-}(\delta\theta_{O}^{C}\delta\theta_{O+1}^{C} -\delta\theta_{O}^{F}\delta\theta_{O+1}^{F}).
\end{split}
\end{equation}
See details of simplification between Eqs.~\eqref{eqS:DW_withK} and \eqref{eqS:DW_explicit} in Sec. 2.2.3. If we consider the case of $I>O$, the superscript of $k_{i}^{\pm}$ in Eq. \eqref{eqS:DW_explicit} needs to be reversed, which shows the loading path dependency. 

We first note there is a prefactor, $P=\prod_{i=I+1}^{O-1}\frac{k_{i}^{+}}{k_{i}^{-}}$ in front of the term of $k_{I}^{+}(\delta\theta_{I}^{C}\delta\theta_{I+1}^{C} -\delta\theta_{I}^{F}\delta\theta_{I+1}^{F})$. Interestingly, this prefactor becomes nonlocal (i.e., a function of many coupling constants $k_i^p$ and $k_i^a$) as a result of the non-reciprocal couplings $k_{i}^{a}$. If $k_{i}^{a}=0$, the entire prefactor becomes 1 and Eq.~\eqref{eqS:DW_explicit} is reduced to the same expression of elastic energy. If $k_{i}^{a}\neq0$, this prefactor contains all stiffness components, which makes that the derivative of the work against $k_{i}$, $\frac{\partial(\Delta W)}{\partial k_{i}}$, is not only determined by $\delta\theta_{i}$ and $\delta\theta_{i+1}$ but also other $k_{i}$.

\subsubsection{Multiple inputs and outputs}
We next consider the case of multiple inputs and outputs. We notice that the work (Eq.~\eqref{eqS:DW_explicit}) only depends on angular deflections of input unit $I$, output unit $O$, and their nearest neighbor $I+1$, $O-1$ and $O+1$ when there is a single input and output. In other words, the work function is local in terms of angular deflections, so fixing a single input and then nudging a single output can be treated as independent loading. If we intend to apply multiple inputs and nudge several outputs, we assume the total work is equivalent to applying a single input and nudging every output sequentially, back to the initial state (no units are fixed), then applying the next input and again nudging every output sequentially. The total work with multiple inputs and outputs is, therefore, the sum of the work with a single input and a single output. For example, if all indices of the inputs are smaller than those of the outputs, the work difference between clamped and free states reads
\begin{equation}
\begin{split}
    \Delta W
    = & -\frac{1}{2}\sum_{i\in\mathcal{I}} P\, k_{I_{i}}^{+} (\delta\theta_{I_{i}}^{C}\delta\theta_{I_{i}+1}^{C} -\delta\theta_{I_{i}}^{F}\delta\theta_{I_{i}+1}^{F})\\
    & -\frac{1}{2}\sum_{i\in\mathcal{O}} \{
    k_{O_{i}}^{o}[(\delta\theta_{O_{i}}^{C})^2 - (\delta\theta_{O_{i}}^{F})^2]
    + k_{O_{i}-1}^{+}(\delta\theta_{O_{i}-1}^{C}\delta\theta_{O_{i}}^{C} -\delta\theta_{O_{i}-1}^{F} \theta_{O_{i}}^{F})
    + k_{O_{i}}^{-}(\delta\theta_{O_{i}}^{C}\delta\theta_{O_{i}+1}^{C} -\delta\theta_{O_{i}}^{F}\delta\theta_{O_{i}+1}^{F}) \} \\
    = & -\frac{1}{2}\sum_{i\in\mathcal{I}}P\, (k_{i}^{p}+k_{i}^{a})(\delta\theta_{I_{i}}^{C}\delta\theta_{I_{i}+1}^{C} -\delta\theta_{I_{i}}^{F}\delta\theta_{I_{i}+1}^{F}) \\
    & -\sum_{i\in\mathcal{O},i\neq O_{1}}\left \{\frac{1}{2}k_{O_{i}}^{o}[(\delta\theta_{O_{i}}^{C})^2 - (\delta\theta_{O_{i}}^{F})^2]
      +k_{i}^{p}(\delta\theta_{O_{i}-1}^{C}\delta\theta_{O_{i}}^{C} -\delta\theta_{O_{i}-1}^{F} \theta_{O_{i}}^{F})\right \}\\
    & -\frac{1}{2}\left \{ k_{O_{1}}^{o}[(\delta\theta_{O_{1}}^{C})^2 - (\delta\theta_{O_{1}}^{F})^2] 
      + (k_{O_{1}}^{p}+k_{O_{1}}^{a})(\delta\theta_{O_{1}-1}^{C}\delta\theta_{O_{1}}^{C} -\delta\theta_{O_{1}-1}^{F} \theta_{O_{1}}^{F}) \right \}.
\end{split}
\end{equation}
Here, $\mathcal{I}$ and $\mathcal{O}$ are the sets of input and output indices. $I_{i}$ and $O_{i}$ are the $i^{th}$ element of $\mathcal{I}$ and $\mathcal{O}$. If all indices of the inputs are bigger than those of the outputs, the superscript of $k_{i}^{\pm}$ is reversed. If we consider the more general case that the indices of the inputs are not necessarily smaller than those of the outputs, the work can be written as
\begin{equation}
\label{eq:DW_final}
\begin{split}
    \Delta W
    = & -\frac{1}{2}\sum_{i\in\mathcal{I}}P(k_{i}^{p}+\alpha_{i}k_{i}^{a})(\delta\theta_{I_{i}}^{C}\delta\theta_{I_{i}+1}^{C} -\delta\theta_{I_{i}}^{F}\delta\theta_{I_{i}+1}^{F}) \\
    & -\sum_{i\in\mathcal{O},i\neq O_{1}}\{\frac{1}{2}k_{O_{i}}^{o}[(\delta\theta_{O_{i}}^{C})^2 - (\delta\theta_{O_{i}}^{F})^2]+k_{i}^{p}(\delta\theta_{O_{i}-1}^{C}\delta\theta_{O_{i}}^{C} -\delta\theta_{O_{i}-1}^{F} \theta_{O_{i}}^{F}) \}\\
    & -\frac{1}{2}\left \{ k_{O_{1}}^{o}[(\delta\theta_{O_{1}}^{C})^2 - (\delta\theta_{O_{1}}^{F})^2] 
      + (k_{O_{1}}^{p}+\alpha_{O_{1}}k_{O_{1}}^{a})(\delta\theta_{O_{1}-1}^{C}\delta\theta_{O_{1}}^{C} -\delta\theta_{O_{1}-1}^{F} \theta_{O_{1}}^{F}) \right \}.
\end{split}
\end{equation}
Here, $\alpha_i=\mathrm{sgn}(i-I)$ for $i\neq I$, or $\alpha_i=\mathrm{sgn}(O-I)$ for $i=I$, which indicates the direction of the loading path between unit $i$ or output unit $O$ and an input unit $I$. 

\subsubsection{Details of simplifying Eq.~\eqref{eqS:DW_withK}}
\label{SecS:simplifyDW}
The second and third terms in Eq.~\eqref{eqS:DW_withK} can be simplified as
\begin{equation}
\label{eqS:term1,2}
\begin{split}
    & k_{O-1}^{+}(K^L)^{-1}_{O-1, I+1}k_{I}^{+}\delta\bar\theta_{I}\delta\theta_O
      +\frac{1}{2}k_{O-1}^{+}(K^L)^{-1}_{O-1,O-1}k_{O-1}^{-}(\delta\theta_{O})^{2}\\
    & = \frac{1}{2}k_{O-1}^{+}\left[(K^L)^{-1}_{O-1,I+1}\tau^L_{I+1}+(K^L)^{-1}_{O-1,O-1}\tau^L_{O-1}\right]\delta\theta_{O}
      + \frac{1}{2}k_{O-1}^{+}(K^L)^{-1}_{O-1,I+1}\tau^L_{I+1}\delta\theta_{O}\\
    & = \frac{1}{2}k_{O-1}^{+}\left[\sum_{i=I+1}^{O-1}(K^L)^{-1}_{O-1,i}\tau^L_{i}\right]\delta\theta_{O}
      + \frac{1}{2}k_{O-1}^{+}(K^L)^{-1}_{O-1,I+1}\tau^L_{I+1}\delta\theta_{O}\\
    & = -\frac{1}{2}k_{O-1}^{+}\delta\theta_{O-1}\delta\theta_{O}
      + \frac{1}{2}k_{O-1}^{+}(K^L)^{-1}_{O-1,I+1}k_{I}^{+}\delta\bar\theta_I\delta\theta_{O}.
\end{split}
\end{equation}
Next, we simplify the last term as follows:
\begin{equation}
\label{eqS:k_{O-1}+}
\begin{split}
    k_{O-1}^{+}(K^L)^{-1}_{O-1,I+1}k_{I}^{+}\delta\bar\theta_I\delta\theta_{O}\bigg|_{\delta\theta_O^F}^{\delta\theta_O^C}
    & = k_{O-1}^{+}(K^L)^{-1}_{O-1,I+1}k_{I}^{+}\delta\bar\theta_I(\delta\theta_{O}^{C}-\delta\theta_{O}^{F}) \\
    & = \frac{k_{O-1}^{+}}{k_{O-1}^{-}}(K^L)^{-1}_{O-1, I+1}\tau_{I+1}^{L}[(\tau_{O-1}^{L})^{C}-(\tau_{O-1}^{L})^{F}].
\end{split}
\end{equation}

\rv{
where we have used Eq.~\eqref{eqS:tau_L}. Next, we have
\begin{equation}
\label{eqS:tau_O-1}
\begin{split}
    (\tau_{O-1}^{L})^{C}-(\tau_{O-1}^{L})^{F}
    & = \frac{(K^L)^{-1}_{I+1, O-1}}{(K^L)^{-1}_{I+1, O-1}}[(\tau_{O-1}^{L})^{C}-(\tau_{O-1}^{L})^{F}]\\
    & = \frac{1}{(K^L)^{-1}_{I+1, O-1}}\left \{(K^L)^{-1}_{I+1, I+1}[(\tau_{I+1}^{L})^{C}-(\tau_{I+1}^{L})^{F}]+(K^L)^{-1}_{I+1, O-1}[(\tau_{O-1}^{L})^{C}-(\tau_{O-1}^{L})^{F}]\right \}\\
    & = \frac{1}{(K^L)^{-1}_{I+1, O-1}}\sum_{i=I+1}^{O-1}(K^L)^{-1}_{I+1,i}[(\tau_{i}^{L})^{C}-(\tau_{i}^{L})^{F}]\\
    & = -\frac{1}{(K^L)^{-1}_{I+1, O-1}}(\delta\theta_{I+1}^{C}-\delta\theta_{I+1}^{F}),
\end{split}
\end{equation}
where we use the fact that $(\tau_{I+1}^{L})^{C}=(\tau_{I+1}^{L})^{F}=k_{I}^{+}\delta\bar\theta_{I}$ (Eq.~\eqref{eqS:tau_L}).

Substituting Eq.~\eqref{eqS:tau_O-1} to Eq.~\eqref{eqS:k_{O-1}+} , we obtain the following expression and simplify it as
\begin{equation}
\label{eqS:final_term2}
\begin{split}
    k_{O-1}^{+}(K^L)^{-1}_{O-1,I+1}k_{I}^{+}\delta\bar\theta_I\delta\theta_{O}\bigg|_{\delta\theta_O^F}^{\delta\theta_O^C}
    & = -\frac{k_{O-1}^{+}}{k_{O-1}^{-}}\frac{(K^L)^{-1}_{O-1, I+1}}{(K^L)^{-1}_{I+1, O-1}}\tau_{I+1}^{L}(\delta\theta_{I+1}^{C}-\delta\theta_{I+1}^{F})\\
    & = -\frac{(K^L)^{-1}_{O-1, I+1}}{(K^L)^{-1}_{I+1, O-1}}\frac{k_{O-1}^{+}}{k_{O-1}^{-}}k_{I}^{+}(\delta\theta_{I}^{C}\delta\theta_{I+1}^{C}-\delta\theta_{I}^{F}\delta\theta_{I+1}^{F})\\
    & = -\left(\prod_{i=I+1}^{O-2}\frac{k_{i}^{+}}{k_{i}^{-}}\right)\frac{k_{O-1}^{+}}{k_{O-1}^{-}}k_{I}^{+}(\delta\theta_{I}^{C}\delta\theta_{I+1}^{C}-\delta\theta_{I}^{F}\delta\theta_{I+1}^{F})\\
    & = -\left(\prod_{i=I+1}^{O-1}\frac{k_{i}^{+}}{k_{i}^{-}}\right)k_{I}^{+}(\delta\theta_{I}^{C}\delta\theta_{I+1}^{C}-\delta\theta_{I}^{F}\delta\theta_{I+1}^{F}).
\end{split}
\end{equation}
Here, we use the fact that $\delta\theta_{I}^{F}=\delta\theta_{I}^{C}=\delta\bar\theta_{I}$, 
\begin{equation}
    (K^{L})^{-1}_{I+1,O-1}=\frac{(-1)^{I+O}}{\mathrm{det}(K^L)}\prod_{i=I+1}^{O-2}k_{i}^{-},
\end{equation}
and
\begin{equation}
    (K^{L})^{-1}_{O-1,I+1}=\frac{(-1)^{I+O}}{\mathrm{det}(K^L)}\prod_{i=I+1}^{O-2}k_{i}^{+},
\end{equation}
where $\mathrm{det}(K^{L})$ refers to the determinant of $K^{L}$.

The last term in Eq.~\eqref{eqS:DW_withK} can be simplified as
\begin{equation}
\label{eqS:term3}
\begin{split}
     \frac{1}{2}k_{O}^{-}(K^R)^{-1}_{O+1,O+1}k_{O}^{+}(\delta\theta_{O})^{2}
    = & \frac{1}{2}k_{O}^{-}(K^R)^{-1}_{O+1,O+1}\tau_{O+1}\delta\theta_{O}\\
    = & \frac{1}{2}k_{O}^{-}\left[\sum_{i=O+1}^{N}(K^R)^{-1}_{O+1,i}\tau_{i}\right]\delta\theta_{O}\\
    = & -\frac{1}{2}k_{O}^{-}\delta\theta_{O}\delta\theta_{O+1}.
\end{split}
\end{equation}
}
Eventually, we obtain the explicit expression of work (Eq.~\eqref{eqS:DW_explicit}) by substituting Eqs.~\eqref{eqS:term1,2}, \eqref{eqS:final_term2} and \eqref{eqS:term3} into Eq.~\eqref{eqS:DW_withK}.

\subsection{\rv{Path-dependent}~contrastive learning rule}
Now that we have expressed the work difference between the clamped and free states, how can we derive a contrastive learning rule? A contrastive learning rule must be local, translation invariant and needs to lead to a decrease of the cost function $\psi$ during learning. If our learning rule is successful, a decrease of the cost function $\Delta\psi$ should also lead to a decrease of the work difference $\Delta W$, i.e., the free response will approach the clamped response. Therefore, we will construct a cost function that retains the main features of $\Delta W$, yet is local and translation invariant. 

To this end, we introduce
\begin{equation}
\label{eqS:psi}
\Delta\psi = \psi^C-\psi^F=
-\frac{1}{2}\sum_{i=1}^{N}k_{i}^{o}[(\delta\theta_{O_{i}}^{C})^2 - (\delta\theta_{O_{i}}^{F})^2] 
- \sum_{i=1}^{N-1}(k_{i}^{p}+\alpha_{i}k_{i}^{a})(\delta\theta_{i}^{C}\delta\theta_{i+1}^{C} -\delta\theta_{i}^{F}\delta\theta_{i+1}^{F}),
\end{equation}
which corresponds to Eq. (6) of the Main Text. Here, $\alpha_i=\mathrm{sgn}(i-I)$ for $i\neq I$, or $\alpha_i=\mathrm{sgn}(O-I)$ for $i=I$, which indicates the direction of the loading path between unit $i$ or output unit $O$ and an input unit $I$. Note that $I$ and $O$ can be any one of the input and output unit indices. \rv{The stiffness $k_{i}^{a}$ to which $\alpha$ applies sets the magnitude of the anti-symmetric torque. This torque changes sign depending on the direction of actuation.} If $i > I$, i.e., the $i^\textrm{th}$ unit is on the right side of the input $I$, the loading path goes from left to right, $\alpha_i=1$ and the contribution to $\Delta\psi$ by $k_{i}^{a}$ is positive. In contrast, if $i < I$, i.e., the $i^\textrm{th}$ unit is on the left side of the input $I$, the loading path goes backward from right to left, $\alpha_i=-1$ and the contribution to $\Delta\psi$ by $k_{i}^{a}$ is negative. If $i=I$, the contribution of $k_{i}^{a}$ is given by the loading path between input and output units.

In contrast to Eq.~\eqref{eq:DW_final}, $\psi$ is local ($P=1$) and translation invariant (the sum runs over all indices instead of only the output nodes). Note that the additional terms of this sum will not affect the minimization: the contribution of $k_{i}^{a}$ if $i\neq O_{1}$ is canceled and the contribution of $k_{i}^{o}$ if $i\notin\mathcal{O}$ is zero (see Eq.~\eqref{eq:DW_final}). As a result, $\psi$ can be used to conduct any learning task. The key feature of the cost function $\Delta\psi$ in contrast with earlier contrastive learning schemes is that it is path dependent, a crucial aspect of systems with non-reciprocal forces. We substitute Eq.~\eqref{eqS:psi} into Eq.~(2), and then we obtain the explicit local learning rules for our non-reciprocal system as shown in Eqs.~(4), (5) and (7).

\section{Learning space evaluation}
\label{SecS:Learning space evaluation}
We now evaluate the learning space of all used systems by counting the number of degrees of freedom and constraints. Here, the degrees of freedom include the angles and the stiffnesses, and the constraints include the torque balance and angle constraints. A feasible learning solution exists only if the number of constraints is at most equal to the number of degrees of freedom. We analyze the system with $N$ units and first derive the bounds for the single target learning and then for multiple target learning.

\subsection{Single target}
\label{SecS:Learning space evaluation-single target}
Assuming a single target consists of $N_{I}$ input units and $N_{O}$ output units, there are $N-N_{I}$ equations of torque balance ($\tau_{i}=0,\ i \notin \mathcal{I}$) and $N_{I}+N_{O}$ equations of angle constraints ($\delta\theta_{i}=\text{const},\ i \in (\mathcal{I}\cup\mathcal{O})$). So the number of total constraints is $N+N_{O}$. Here, $\mathcal{I}$ and $\mathcal{O}$ are the sets of input and output indices. We then calculate the number of degrees of freedom in different system configurations.
 
For the $p$ configuration (Eq. (1) with $k_{i}^{a}=0$), there are $2N-1$ independent stiffness parameters and $N$ angular deflections. The condition of obtaining a solution is
\begin{equation}
\label{eqS:condition_p_single}
(3N-1)-(N+N_{O})=2N-N_{O}-1 \geq 0.
\end{equation}
For the $a$ configuration (Eq. (1) with $k_{i}^{a}\neq 0$), there are $3N-2$ independent stiffness parameters and $N$ angular deflections. The condition is
\begin{equation}
\label{eqS:condition_a_single}
(4N-2)-(N+N_{O})=3N-N_{O}-2 \geq 0.
\end{equation}
For the $pp$ configuration (Eq.~(M13) with $k_{i}^{a}=k_{i}^{aa}=0$), there are $3N-3$ independent stiffness parameters and $N$ angular deflections. The condition is 
\begin{equation}
\label{eqS:condition_pp_single}
(4N-3)-(N+N_{O})=3N-N_{O}-3 \geq 0.
\end{equation}
For the $aa$ configuration (Eq.~(M13) with $k_{i}^{a}\neq 0\ \text{and}\ k_{i}^{aa}\neq 0$), there are $5N-6$ independent stiffness parameters and $N$ angular deflections. The condition is
\begin{equation}
\label{eqS:condition_aa_single}
(6N-6)-(N+N_{O})=5N-N_{O}-6 \geq 0.
\end{equation}

The above relations are shown in Fig.~\ref{figS:learning space}a and Table.~\ref{tabS:learning space} as well. The $aa$ system has the largest learning space, which means the best performance for the single target learning, and what follows in order are $a$, $pp$ and $p$ systems. It is consistent with the results in Fig.~\ref{figS:single target learning}.

\subsection{Multiple targets}
\label{SecS:Learning space evaluation-multiple targets}
Different from single target learning, a system learns $N_{T}$ targets simultaneously and each target consists of $N_{I}$ input units and $N_{O}$ output units. Hence, there are are $N_{T}(N-N_{I})$ equations of torque balance and $N_{T}(N_{I}+N_{O})$ equations of angle constraints. The number of degrees of freedom is the same as above.

For the $p$ configuration, the condition of a feasible solution is
\begin{equation}
(3N-1)-N_{T}(N+N_{O})\geq 0.
\end{equation}
For the $a$ configuration, the condition is
\begin{equation}
(4N-2)-N_{T}(N+N_{O})\geq 0.
\end{equation}
For the $pp$ configuration, the condition is
\begin{equation}
(4N-3)-N_{T}(N+N_{O})\geq 0.
\end{equation}
For the $aa$ configuration, the condition is
\begin{equation}
(6N-6)-N_{T}(N+N_{O})\geq 0.
\end{equation}

\begin{figure}[htbp!]
    \centering
    \includegraphics[width=1\linewidth]{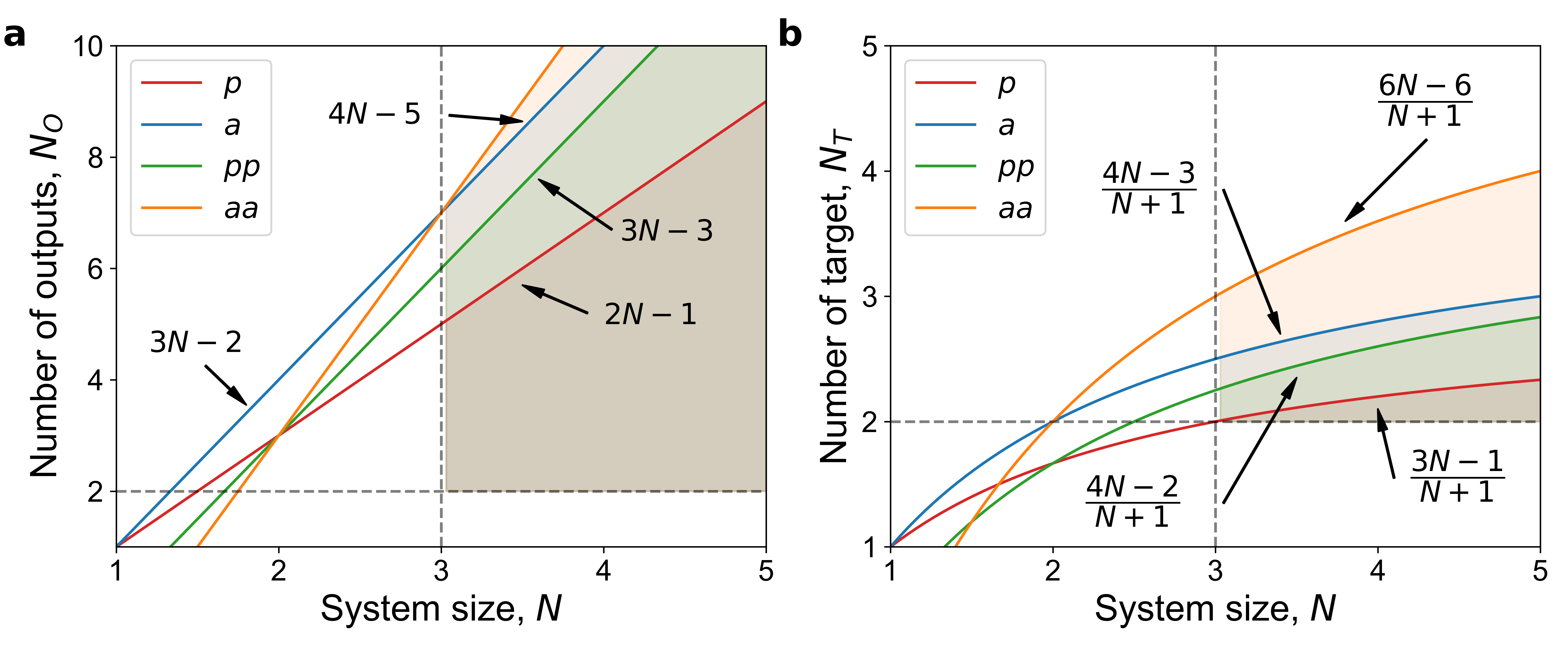}
    \caption{\textbf{Learning space of different system configurations}. The shaded regions represent where the conditions of a feasible solution are satisfied, and the area is equivalent to the learning space volume. Here, it displays the learning space when $N>3$ as examples. 
    {\bf a,} Single target learning. $N_{O}$ is the output number. We take $N_{O}>2$ as an example. The shaded regions means where Eqs.~\eqref{eqS:condition_p_single},~\eqref{eqS:condition_a_single},~\eqref{eqS:condition_pp_single} 
    and~\eqref{eqS:condition_aa_single} meet if $N_{O}>2$, respectively. 
    {\bf b,}  Multi-target learning. $N_{T}$ is the target number. We take $N_{T}>2$ as an example. The shaded regions indicate where Eqs.~\eqref{eqS:condition_p_multi},~\eqref{eqS:condition_a_multi},~\eqref{eqS:condition_pp_multi} 
    and~\eqref{eqS:condition_aa_multi} meet if $N_{T}>2$, respectively.}
    \label{figS:learning space}
\end{figure}

If we choose $N_{I}=N_{O}=1$ as the same consideration in Fig. 2(d), the above conditions are
\begin{equation}
\label{eqS:condition_p_multi}
    N_{T}<\frac{3N-1}{N+1}, \text{ for the $p$ configuration},
\end{equation}
\begin{equation}
\label{eqS:condition_a_multi}
    N_{T}<\frac{4N-2}{N+1}, \text{ for the $a$ configuration}.
\end{equation}
\begin{equation}
\label{eqS:condition_pp_multi}
    N_{T}<\frac{4N-3}{N+1}, \text{ for the $pp$ configuration},
\end{equation}
\begin{equation}
\label{eqS:condition_aa_multi}
    N_{T}<\frac{6N-6}{N+1}, \text{ for the $aa$ configuration}.
\end{equation}
which is shown in Fig.~\ref{figS:learning space}b and Table~\ref{tabS:learning space}. Likewise, the $aa$ configuration has the largest learning space in the case of multiple target learning, and what follows in order are $a$, $pp$ and $p$ configurations. It is also consistent with the results in Fig.~2d. However, we note that the above evaluation ignores the constraints according to the Maxwell-Betti theorem.

\begin{table}[h]
   \caption{\textbf{The evaluation of the learning space of different system configurations.}}
   \label{tabS:learning space}
   \centering
   \begin{tabular}{ccccc}
   \toprule\toprule
   \textbf{Configuration} & \textbf{$p$} & \textbf{$a$} & \textbf{$pp$} & \textbf{$aa$} \\ 
   \midrule
   Single target    & $2N-1$ & $3N-2$ & $3N-3$ & $5N-6$ \\
   \\
   Multiple targets & $\dfrac{3N-1}{N+1}$ & $\dfrac{4N-2}{N+1}$ & $\dfrac{4N-3}{N+1}$ & $\dfrac{6N-6}{N+1}$\\
   \bottomrule
   \end{tabular}
\end{table}

\section{Single target learning \rv{at small scales}}
\label{SecS:Single target learning}
\begin{figure}[htbp!]
    \centering
    \includegraphics[width=1\linewidth]{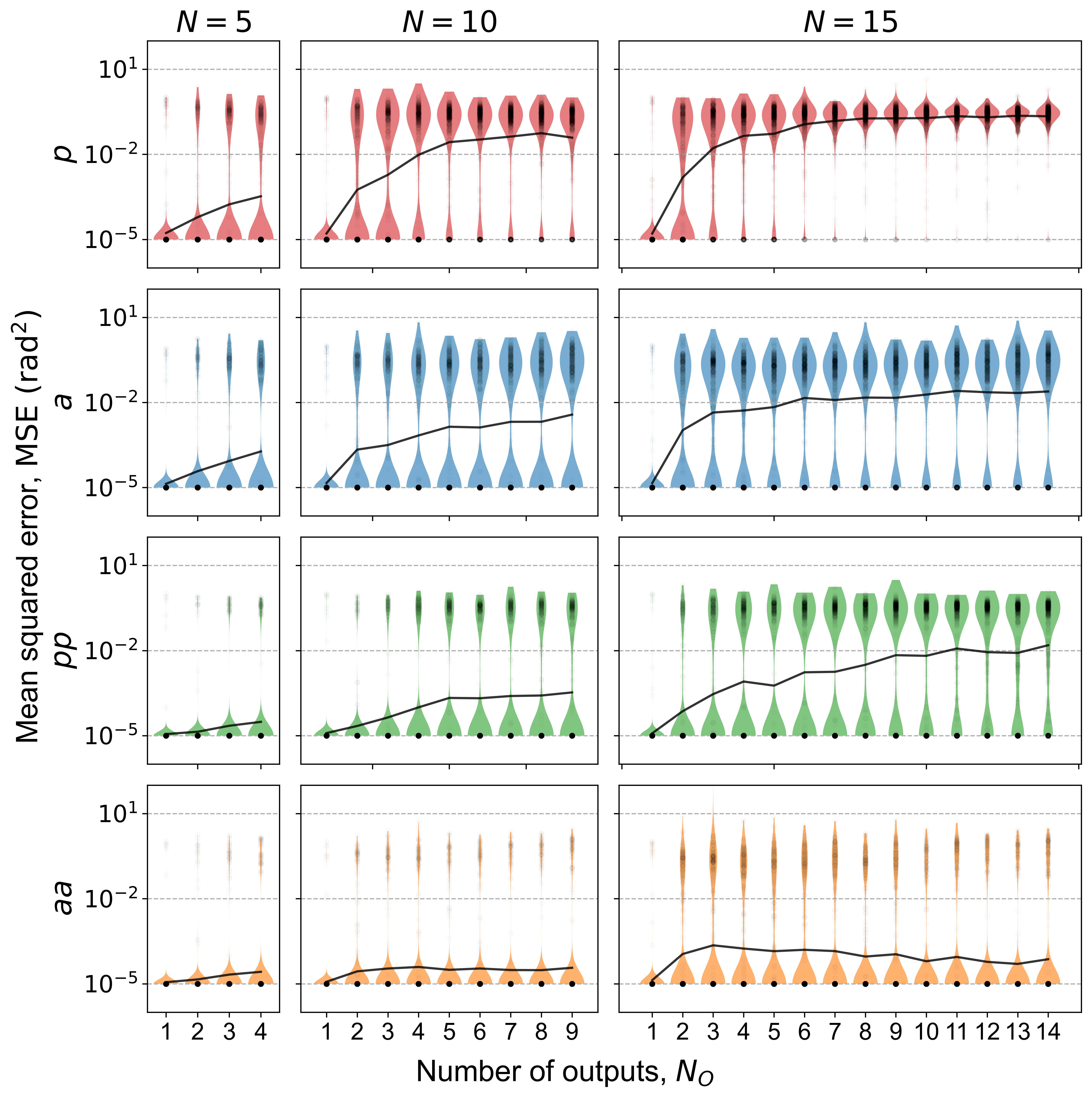}
    \caption{\textbf{Simulation results of learning single target with (non)reciprocal, and next nearest neighbor interactions ($p$, $a$, $pp$ and $aa$ configurations)}. Systems of $N$ = 5, 10 and 15 are simulated, and the number of output units $N_{O}$ varied from 1 to $N-1$. The black semi-transparent dots are the MSE of each simulation and each column consists of 500 simulations. The initial parameters are $k_{i}^{o}=0.1$, $k_{i}^{p}=0.01$, $k_{i}^{a}=0$, $k_{i}^{pp}=0.01$ and $k_{i}^{aa}=0$. The solid line is the average MSE. The cut-off of the MSE is arbitrarily chosen to be $10^{-5}\ \mathrm{rad}^2$.}
    \label{figS:single target learning}
\end{figure}

We simulate a system with $N$ units to learn a single target with multiple outputs and compare the learning performance of different system configurations. Here, a single target consists of a single randomly selected input unit and $N_{O}$ randomly selected output units. We compare four system configurations: $p$, $a$, $pp$ and $aa$ for $N$=5, 10, and 15, and vary $N_{O}$ from 1 to $N-1$ (Fig.~\ref{figS:single target learning}). 

\rv{
For the same system configuration and the same $N_{O}$, the MSE increases as the system size $N$ grows. This is due to the decay of deformation: the effect of a deformation of one node on its neighbors' deformations decreases exponentially with distance. As a result, as the system size grows, the average distance between the input and outputs also increases, and learning converges more slowly. We discuss deformation decay in more detail in Sec.~5.1.

Surprisingly, in our systems with $N=10$ and 15, the MSE initially rises, then saturates, or even reduces in the $aa$ system as $N_{O}$ increases. This is because having more outputs suppresses the deformation decay effect (see Sec.~5.1). Overall, the simplest $p$ configuration performs the worst. Introducing non-reciprocal interactions $k_{i}^{a}$ leads to a lower MSE in the $a$ configuration, which lowers even further by introducing second nearest-neighbor interactions $k_{i}^{pp}$ and $k_{i}^{aa}$ in the $pp$ and $aa$ configurations. Among these four configurations, the $aa$ configuration performs best with the lowest MSE. This outcome is expected as adding more learning degrees of freedom expands the learning space (see Sec.~\ref{SecS:Learning space evaluation}).
}

\section{\rv{Single target learning at large scales}}
In experiments and Sec.~\ref{SecS:Single target learning}, our metamaterials are limited to approximately 15 units. Here, we ask how contrastive learning performs at larger scales. In simulations, we notice that the MSE increases with the system size $N$(Fig.~\ref{figS:single target learning}). We hypothesize that learning becomes more difficult when the input and output are far apart, owing to the aforementioned deformation decay effect. To test our hypothesis, we first analyze the deformation decay effect mathematically. Next, we assess its impact on learning performance in the simplest case of a single input and output. Building on these results, we finally investigate how the learning scheme performs at larger scales.

\begin{figure}[htbp!]
    \centering
    \includegraphics[width=1\linewidth]{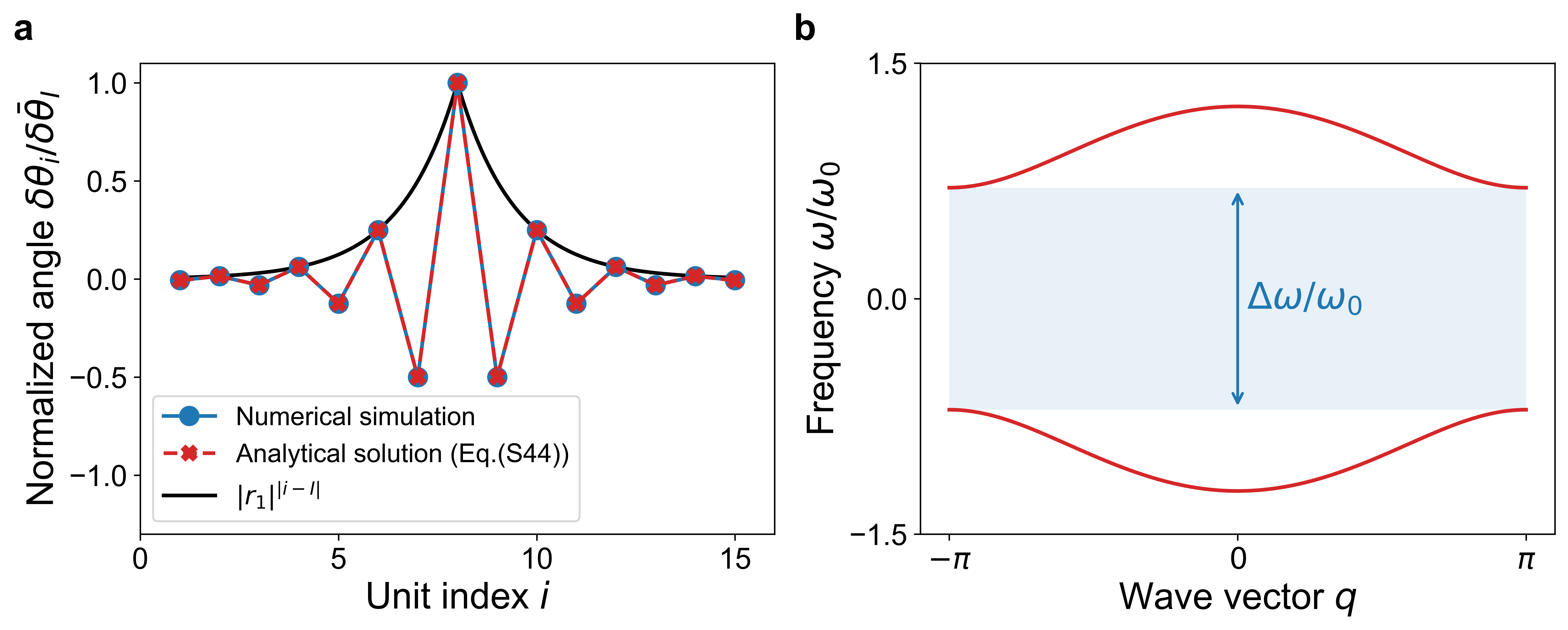}
    \caption{\rv{
    \textbf{Analysis of the deformation decay effect.}
    \textbf{a,} The discrete field of the angular deflection. Here, we consider a system with $N$ = 15, $k^{o}=2$, $k^{p}=0.5$, and an input angle $\delta\bar\theta_{I}$ applies on the $\text{8}^{\text{th}}$ unit. It is equivalent to a passive chain with symmetric nearest-neighbor interactions. The solid line shows the absolute value of the angular deflection decays exponentially from the input.
    \textbf{b,} Rescaled frequency $\omega/\omega_{0}$ versus wave vector $q$ from the Fourier analysis. The blue shaded area denotes the band-gap $\Delta\omega$. Here, we use $k^{o}=2$, $k^{p}=0.5$, $m=1$ and $\delta\theta=\pi/16$.}}
    \label{figS:deformation_decay}
\end{figure}

\rv{
\subsection{Deformation decay effect}
\label{SecS:Deformation_decay_effect}
To assess how the deformation decays in our systems, we consider a passive chain with symmetric nearest-neighbor interactions, such that $k_{i}^{o}=k^{o}$, $k_{i}^{p}=k^{p}$ and $k_{i}^{a}=k^{e}=0$. The torque on unit $i$ is then given by
\begin{equation}
\label{eqS:tau_p}
    \tau_{i}=-k^{o}\delta\theta_{i}-k^{p}(\delta\theta_{i-1}+\delta\theta_{i+1}).
\end{equation}

We are interested in the equilibrium configurations of our chain. To this end, we aim to solve $\tau_i = 0$ for all $i$. We take the ansatz $\delta\theta_{i}=Ar^{i}$, where $r$ is a constant to be determined and $A$ is a constant factor determined by the boundary conditions. By substituting this ansatz into Eq.~\eqref{eqS:tau_p}, we obtain a characteristic equation:
\begin{equation}
    k^{p}r^{2}+k^{o}r+k^{p}=0.
\end{equation}
There are two roots:
\begin{equation}
    r_{\pm}=\frac{-k^{o}\pm\sqrt{(k^{o})^2-4(k^{p})^2}}{2k^{p}}.
\end{equation}
Applying a constant angular deflection $\delta\bar\theta_{I}$ on the $I^\text{th}$ unit is equivalent to setting the boundary condition $\delta \theta_I = \delta\bar\theta_{I}$. We ensure the system is stable, i.e., $|k^{o}/k^{p}|>2$, see Sec.~\ref{SecS:Stability analysis}. The solution to Eq.~\eqref{eqS:tau_p} then reads as
\begin{equation}
\label{eqS:theta_decay_i}
    \delta\theta_{i}=\delta\bar\theta_{I}r_{1}^{i-I}.
\end{equation}
where $r_{1}$ is the smallest root and satisfies $|r_{1}|<1$. The latter follows from $|k^{o}/k^{p}|>2$ and $r_{+}r_{-}=1$ (Vieta's formulas). Thus, we find that the deformations decay exponentially as $\delta \theta_i \propto \exp((i-I) \ln{r_{1}})$. 

To check our derivation, we consider a system with $N$ = 15, $k^{o}=2$, $k^{p}=0.5$ and an input angle $\delta\bar\theta_{I}$ on the $\text{8}^{\text{th}}$ unit. As shown in Fig.~\ref{figS:deformation_decay}a, the angular deflections decay exponentially from the input unit, and Eq.~\eqref{eqS:theta_decay_i} fits the numerical results well. We note that the staggered angular deflection arises when taking $k^{o}/k^{p}>0$ and $r_1<0$, while no such staggering occurs when $k^{o}/k^{p}<0$ and $r_1>0$.

In addition, we take the continuum limit: $\delta\theta i\rightarrow \delta\theta(x)$. We do a Taylor series around $i=n$: $\delta\theta_{i-1}=\delta\theta(x)-a\frac{\mathrm{d}(\delta\theta)}{\mathrm{d}x}+\frac{1}{2}a^{2}\frac{\mathrm{d}^{2}(\delta\theta)}{\mathrm{d}x^{2}}$ and $\delta\theta_{i+1}=\delta\theta(x)+a\frac{\mathrm{d}(\delta\theta)}{\mathrm{d}x}+\frac{1}{2}a^{2}\frac{\mathrm{d}^{2}(\delta\theta)}{\mathrm{d}x^{2}}$, where $a=\mathrm{d}x$. Next, we perform a change of units $x\rightarrow x/a$ and obtain
\begin{equation}
    \frac{\mathrm{d}^{2}(\delta\theta)}{\mathrm{d}x^{2}}-\Omega^{2}\delta\theta=0
\end{equation}
where $\Omega^{2}=-(\beta+2)/a^{2}$ and $\beta=k^{o}/k^{p}$. Solutions for the continuum field $\delta\theta(x)$ are linear combinations of $\{e^{\Omega x}, e^{-\Omega x}\}$, and we again find a spatial exponential decay of the deformation.

Finally, we analyze the band structure in our system. We consider an infinite system described by Eq.~\eqref{eqS:tau_p} and take the discrete Fourier transform: $\delta \theta_n = \frac{1}{N}\sum_{q} \delta \hat{\theta}_q \exp(i q n)\exp(-\omega t)$, where $q$ is the wave vector. To avoid confusion, we replace $i$ with $n$ as the unit index and reserve $i$ to represent the imaginary number in this paragraph. Using the orthogonality of discrete Fourier modes, we obtain
\begin{equation}
    \left[\omega^2-D(q)\right]\delta \hat{\theta}_q = 0,
\label{eqS:dispersion_relation}
\end{equation}
where
\begin{equation}
    D(q)=-\omega_{0}^{2}\left(1+2\frac{k^{p}}{k^{o}}\mathrm{cos}(q)\right)
\end{equation}
is a scalar that can be interpreted as the Fourier transform of the dynamical matrix. Here, $\omega_{0}=\sqrt{k^{o}/m}$ denotes the characteristic frequency of the structure, where $m$ is the mass of each unit. We plot the dispersion relation in Fig.~\ref{figS:deformation_decay}b. We note that there is a band-gap $\Delta\omega$ if $|k^{o}/k^{p}|>2$ around $\omega=0$. This means that there is no real wave vector $q$ that satisfies a static solution. Instead, we can analytically continue $q$ to the imaginary domain as $q\rightarrow  i \kappa$ and solve Eq.~\eqref{eqS:dispersion_relation} for $\omega^2=0$, i.e., we are looking for static evanescent waves. We find the solution $\kappa=\cosh^{-1}(-k^o/2 k^p)$. If we transform back to real space, we find $\delta \theta_n \propto \exp{(\kappa n)}$. This is consistent with the solution we found before, where $\delta \theta_n\propto r_1^{n}$. To see this, we note that $\cosh^{-1}(x) = \ln(x+\sqrt{x^2-1})$ for $x>1$ and
$\cosh^{-1}(-x)=\ln(x+\sqrt{x^2-1}) - \ln(-1)$ for $x>1$ . Using this relation, we find that $\exp(-\kappa n) = (-k^o/2 k^p - \sqrt{(k^o/2 k^p)^2-1})^n = r_{-}^n$ for $k^o/k^p < -2$ and $\exp(-\kappa n) = (-1)^n(k^o/2 k^p - \sqrt{(k^o/2 k^p)^2-1})^n = r_{+}^n$ for $k^o/k^p>2$. Note that the respective root $r_{\pm}$ corresponds to the smallest root $|r_1|<1$. Thus, we find that we get evanescent static waves that decay exponentially.

Overall, this deformation decay effect leads to a slow convergence for a learning task if the input and output units are far apart. Since the angular deflection $\delta\theta$ of the output is small, the rate of updating $\mathrm{d} k_{i}/\mathrm{d}t$ in Eqs. (4), (5), (7) and (M15) are also small. 
% In other words, the distance between the input and output determines the learning convergence. 
To show that the deformation decay affects learning, we next explore how the distance between input and output affects the learning performance. This is directly related to the scaling of our learning scheme as well.
}

\begin{figure}[htbp!]
    \centering
    \includegraphics[width=1\linewidth]{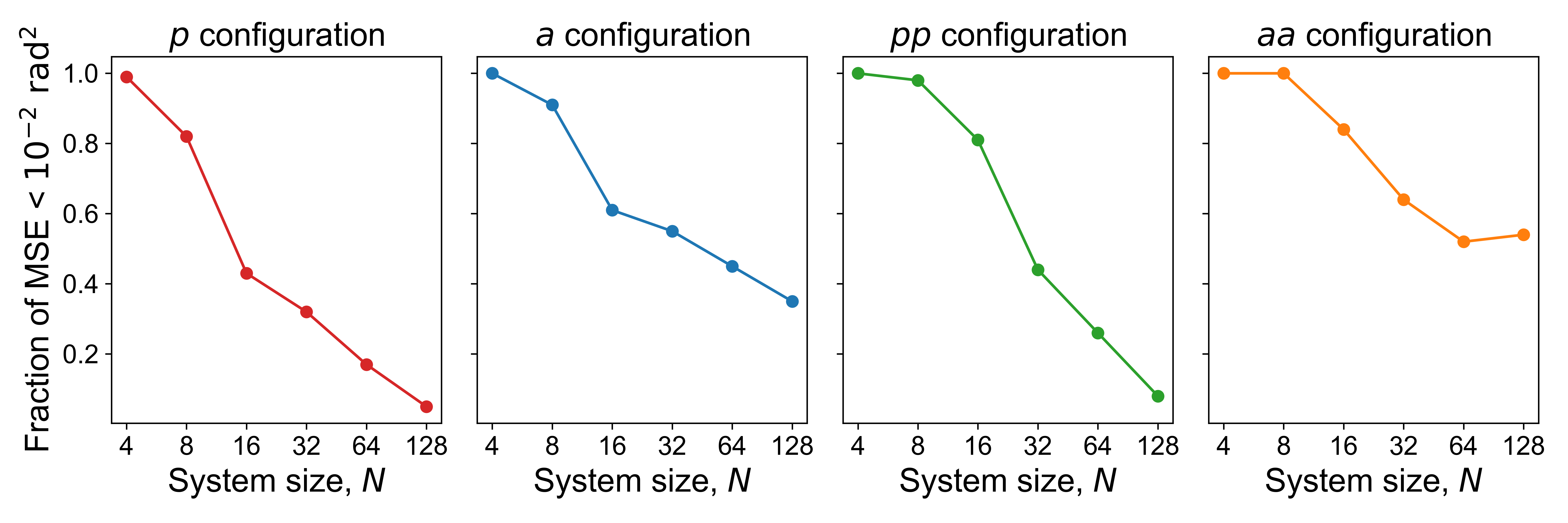}
    \caption{\rv{
    \textbf{The effect of deformation decay on learning.} Considering a system with $N$ units learns a target where a single input is on the unit 1 and a single output is on the unit $N$. The distance between the input and output, i.e., the system size $N$, could be interpreted as the strength of the deformation decay effect. We vary $N$ from 4 to 128 and run 100 simulations for each size. Here, it shows the fraction of simulations with MSE below $10^{-2}\ \mathrm{rad}^2$ for different system configurations (see the Main Text). The initial parameters are $k_{i}^{o}=0.02$, $k_{i}^{p}=0.01$, $k_{i}^{a}=0$, $k_{i}^{pp}=0$ and $k_{i}^{aa}=0$. The learning rate $\gamma=10^{-4}$ and each simulation runs 20000 epochs.}}
    \label{figS:fraction_distance}
\end{figure}

\rv{
\subsection{Single input and output}
We now investigate the effect of deformation decay on the learning performance. We consider a system with $N$ units that learns a target where a single input is on unit 1 and a single output is on the last unit $N$. By varying the system size $N$ and running 100 simulations with randomly chosen input and output angles, Fig.~\ref{figS:fraction_distance} depicts the fraction of successfully learned simulations with respect to the system sizes. The successfully learned simulation refers to one simulation with an MSE below $10^{-2}\ \mathrm{rad}^{2}$ after 20000 epochs. Here, we show all four system configurations mentioned in the Main Text. While increasing the system size $N$ and thus the distance between input and output nodes, the fraction of successfully learned simulations drops. However, the metamaterial can still keep learning more than half of the time in the $aa$ configuration even with $N=128$ units. The fractions in the non-reciprocal configurations ($a$ and $aa$) are generally higher than those in the reciprocal configurations ($p$ and $pp$). These results indicate that the deformation decay indeed limits learning performance, but it can be improved by adding non-reciprocal and longer-range interactions. Crucially, this benchmark is an extreme scenario with a single input and a single output that are far apart. In scenarios where we wish to learn a change-shape, we typically consider multiple outputs: this naturally reduces the distance between inputs and outputs and reduces the distance information has to travel. It is supposed to mitigate the adverse effects of deformation decay on learning. We thereby investigate learning performance in the next section, where multiple outputs are involved.
}

\rv{
\subsection{Single input and multiple outputs}
We now extend the case where our metamaterial learns a target with one input and $N-1$ outputs, and assess the learning performance at large scales. Here, we consider the system with the second nearest-neighbor interaction (Eq.~(M13)) and allow updating for $k_{i}^{a}$ and $k_{i}^{aa}$, which has the best learning performance (Figs.~2d, \ref{figS:single target learning} and \ref{figS:fraction_distance}). Besides investigating the effect of system size, we also take into account the effect of target complexity at large scales.
}

\rv{
\subsubsection{The effect of system size}
\label{SecS:system_size}
Here, our metamaterial learns a target with one input and $N-1$ outputs. The angular deflections of the input and outputs are randomly chosen from a uniform distribution with a range of $[-\delta\bar\theta_{\text{max}}, -\delta\bar\theta_{\text{min}}]\cup[\delta\bar\theta_{\text{min}}, \delta\bar\theta_{\text{max}}]$. We set $\delta\bar\theta_{\text{min}}=20^{\circ}$ and $\delta\bar\theta_{\text{max}}=60^{\circ}$. The reason for setting a $\delta\bar\theta_{\text{min}}$ is to prevent nearly zero deformation that will lead to learning failure. It is because if the desired output is a nearly zero value, the related interaction of this unit would converge to nearly zero as well, and deformation is hard to transport to the next unit then which leads to learning failure. We vary the system size from 8 to 1024 units, run 100 simulations for each system size, and eventually extract the error curves to estimate the learning performance.

The convergence speed $S$ is used as a metric to estimate the learning performance of the model. It is computed as the average slope of the error curves across multiple simulations, given by 
\begin{equation}
    S = \frac{1}{N_{s}}\sum_{i}^{N_{s}}S_{i},
\end{equation}
where $N_{s}$ is total number of simulations and $S_{i}$ is the slope of the linear fit for the $i^{\text{th}}$ run. Specifically, for each individual simulation, the MSE in log scale is fitted to a linear model of the form 
\begin{equation}
    \log(\text{MSE}) = -S_{i}*\text{Epoch}+b_{i}.
\end{equation}
Here, $S_{i}$ represents the slope and $b_{i}$ is the intercept of the fitted line. A higher $S_{i}$ indicates that the learning converges faster. In Extended Data Fig.~3a, the results show that the learning converges more slowly as the system size increases. Despite up to a thousand units, our metamaterial can still learn the target effectively. 

Furthermore, we show the error of each output during learning in Extended Data Fig.~3b. Here, we consider a $aa$ system with $N=1024$ that learns a target where the input unit is the middle unit and all other units are outputs. The input unit is the $512^{\text{th}}$ unit and the input angle is $\delta\theta_{512}=30^{\circ}$. The desired angle is $30^{\circ}$ as well for all outputs. We plot the the squared error, $(\delta\theta_{i}^{C}-\delta\theta_{i}^{F})^2$, for each unit during learning. The results indicate that the learning error decays across the system, leading to successful learning. This suggests that involving multiple units helps maintain the flow of information (deformation) throughout the system, mitigating the typical deformation decay observed when there is only one or a few outputs.

In fact, in this limit case where one learns a target with a single input and $N-1$ outputs, the information can propagate over arbitrarily long distances. The information requires some time to propagate through the materials, and therefore, the convergence speed slows down with the system size increase, yet converges for a large system size (Extended Data Fig.~3a). We attribute this saturation to the constant rate at which information propagates in the system. Therefore, we highlight that the propagation of information during learning is inherently linked to the propagation of deformations within the structure, and making deformations long-range by involving multiple outputs helps to mitigate the deformation decay and improve learning performance. These results underscore the generality and robustness of our contrastive learning protocol, even at large scales.
}

\begin{figure}[htbp!]
    \centering
    \includegraphics[width=1\linewidth]{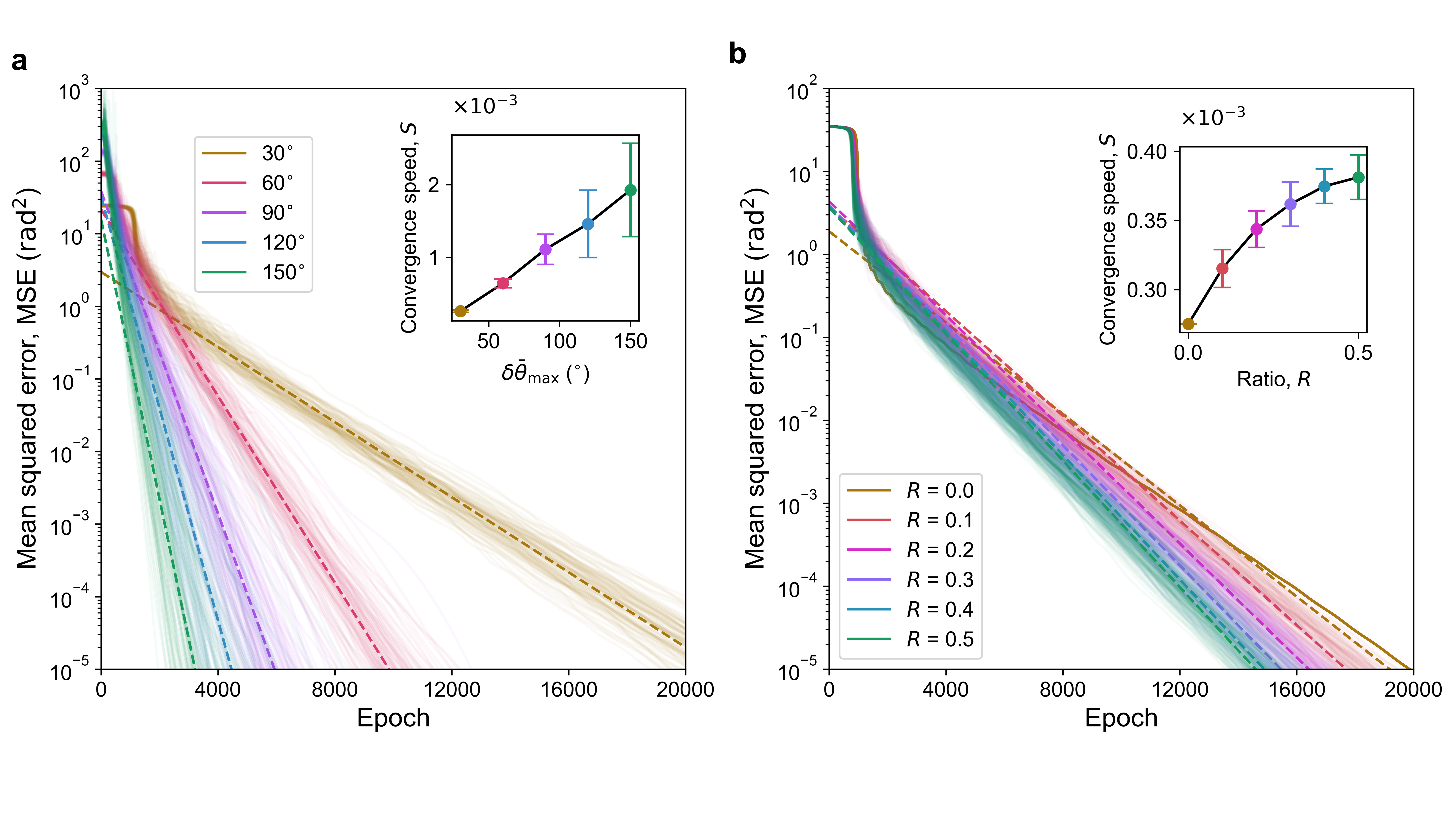}
    \caption{
    \rv{\textbf{The MSE curves of learning with varying target complexity.} We estimate the change of curvature of a desired shape change as the target complexity. Intuitively, a more complex shape change involves more inflection points and has a higher mixture of large and small angular deflections. Therefore, we define two quantities to estimate target complexity: the maximum value of the random desired output angles $\delta\bar\theta_{\mathrm{max}}$ and the ratio of output with opposite sign $R$. Here, we consider a system with $N=128$ units and the second nearest-neighbor interaction ($aa$ configuration, Eq.~(M13)) learns a single target including one input and $N-1$ outputs. We then vary $\delta\bar\theta_{\mathrm{max}}$ and $R$ and run 100 simulations for each value.
    \textbf{a,} The MSE curves with varying $\delta\bar\theta_{\mathrm{max}}$ from $30^{\circ}$ to $150^{\circ}$.  
    \textbf{b,} The MSE curves with varying $R$ from 0.0 to 0.5. The absolute value of all input and output angles is $30^{\circ}$. The dashed lines are the averaged linear fitting of these 100 error curves. These insets also show the convergence speed of the MSE curves $S$ versus  $\delta\bar\theta_{\mathrm{max}}$ and $R$, respectively. See the definition of the convergence speed $S$ in the caption of Extended Data Fig.~3a. Here, the initial parameters are $k_{i}^{o}=0.1$, $k_{i}^{p}=0.01$, $k_{i}^{a}=0$, $k_{i}^{pp}=0$ and $k_{i}^{aa}=0$. The learning rate $\gamma$ is $10^{-4}$.}}
    \label{figS:target_complexity}
\end{figure}

\rv{
\subsubsection{The effect of target complexity}
Here, we define the complexity of a target as the change of curvature of a desired shape change. Intuitively, a more complex shape change involves more inflection points and has a higher mixture of large and small angular deflections. To quantify this target complexity, we use two quantities: (i) The maximum value of the random desired output angles, $\delta\bar\theta_{\mathrm{max}}$, quantifies the range of angular deflections. A higher $\delta\bar\theta_{\mathrm{max}}$ indicates a larger variation among the output angles, thereby a higher mixture of both large and small deflections. (ii) The ratio of output angles with opposite sign, $R$, quantifies the extent of rotational alternation in the target shape. A higher $R$ indicates that a larger proportion of units rotate in opposite directions, corresponding to more inflection points in the desired shape.

In Fig.~\ref{figS:target_complexity}a, we consider a system with $N=128$ units as an example and train it to learn a single target, still including one input and $N-1$ outputs. The angular deflections of the input and outputs are also randomly chosen from a uniform distribution with a range of $[-\delta\bar\theta_{\text{max}}, -\delta\bar\theta_{\text{min}}]\cup[\delta\bar\theta_{\text{min}}, \delta\bar\theta_{\text{max}}]$. Here, we fix $\delta\bar\theta_{\text{min}}=20^{\circ}$ and vary $\delta\bar\theta_{\text{max}}$ from $30^{\circ}$ to $150^{\circ}$. 100 simulations are performed for each different $\delta\bar\theta_{\text{max}}$. The result shows that the error converges more quickly as $\delta\bar\theta_{\text{max}}$ increases. This is not surprising, since larger deformations induce larger updating increments $\frac{\mathrm{d} k_{i}}{\mathrm{d}t}$. In addition, the deformation decay effect is suppressed more since larger deformation propagates further in the system.

In Fig.~\ref{figS:target_complexity}b, we investigate how the number of inflection points in the target affects the learning convergence. To this end, we again simulate a system with $N=128$ units and let it learn a single target including one input and $N-1$ outputs. But we randomly choose $M$ outputs to be positive angles and $N-M-1$ outputs to be negative angles. The absolute values of all input and output angles are $30^{\circ}$. The ratio of positive angles is defined as $R=M/N$, and we vary $R$ from $0.0$ to $0.5$. The results show that the convergence speed of learning slightly decreases as $R$ increases. It indicates that the system learns more slowly when the target shape involves more inflection points.

In addition, we use a system with $N=48$ units, and it successfully learns to morph into the shape of a cat in the experiment (Extended Data Fig.~4). It also proves our metamaterial can learn complex shape changes.
}

\rv{
\section{Stability analysis}
\label{SecS:Stability analysis}
\subsection{Linear stability analysis of a 2-unit system}
\label{SecS:Linear stability analysis}
Because the metamaterials learn static shape changes, we analyze the overdamped linear stability of our system here. This hypothesis is not necessarily valid in experiments as the damping is slightly over the critical damping. Here, we take a 2-unit system as an example of linear stability analysis. Its overdamped dynamical function is
\begin{equation}
\delta\dot{{\Theta}} = -K\delta{{\Theta}}.
\end{equation}
It is equivalent to
\begin{equation}
\label{eqS:2unitsystem_overdamped}
    \binom{\delta\dot{{\theta}}_{1}}{\delta\dot{\theta}_{2}}=
    -\begin{bmatrix}
     k_{1}^{o}   & k_{1}^{p}-k_{1}^{a} \\ 
     k_{1}^{p}+k_{1}^{a} & k_{2}^{o}
    \end{bmatrix}
    \binom{\delta\theta_{1}}{\delta\theta_{2}},
\end{equation}
where $\dot{x}$ denotes the first-order derivative of parameter $x$ with respect to time $t$. Noted that we take $k^{e}=0$ for simplicity, but without loss of generality. 

The analytical solution of Eq.~\eqref{eqS:2unitsystem_overdamped} is
\begin{equation}
\label{eqS:deltatheta(t)_2unit}
\delta\Theta(t)=A_{1}\mathbf{V}_{1}e^{-\lambda_{1}t}+A_{2}\mathbf{V}_{2}e^{-\lambda_{2}t},
\end{equation}
where $A_{i}$ is constant factor, $\lambda_{i}$ and $\mathbf{V}_{i}$ is the $i^{\text{th}}$ eigenvalue and eigenvector of stiffness matrix $K$. The eigenvalues $\lambda_{1,2}$ are
\begin{equation}
    \lambda_{1,2}=\frac{\mathrm{Tr}(K)\pm\sqrt{\Delta}}{2},
\end{equation}
where $\Delta=\mathrm{Tr}(K)^2-4\mathrm{det}(K)$. $\mathrm{Tr}(K)$ and $\mathrm{det}(K)$ are the trace and determinant of $K$, i.e., $\mathrm{Tr}(K)=k_{1}^{o}+k_{2}^{o}$ and $\mathrm{det}(K)=k_{1}^{o}k_{2}^{o}-(k_{1}^{p}-k_{1}^{a})(k_{1}^{p}+k_{1}^{a})$.

The eigenvalues $\lambda_{1,2}$ determine the linear stability of this 2-unit system. In Extended Data Fig.~5a, we plot the stability phase diagram in the space of $\mathrm{Tr}(K)$ and $\mathrm{det}(K)$. In detail, there are five distinct cases:
\begin{enumerate}
    \item $\lambda_{1,2}$ are two complex numbers with a negative real part when $\mathrm{Tr}(K)<0$ and $\Delta<0$. Eq.~\eqref{eqS:deltatheta(t)_2unit} is an oscillation solution with exponentially growing amplitude, indicating that the system is unstable.
    \item $\lambda_{1,2}$ are two complex numbers with a positive real part when $\mathrm{Tr}(K)>0$ and $\Delta<0$. Eq.~\eqref{eqS:deltatheta(t)_2unit} is an oscillation solution with exponentially decaying amplitude, indicating that the system is stable.
    \item $\lambda_{1,2}$ are two real negative numbers when $\mathrm{Tr}(K)<0$, $\mathrm{det}(K)>0$ and $\Delta>0$. The solution given by Eq.~\eqref{eqS:deltatheta(t)_2unit} grows exponentially, and the system is unstable.
    \item $\lambda_{1,2}$ are two real positive numbers when $\mathrm{Tr}(K)>0$, $\mathrm{det}(K)>0$ and $\Delta>0$. The solution given by Eq.~\eqref{eqS:deltatheta(t)_2unit} decays exponentially, and the system is stable.
    \item $\lambda_{1,2}$ are two real numbers, but with opposite signs when $\mathrm{det}(K)<0$. Here, the solution given by Eq.~\eqref{eqS:deltatheta(t)_2unit} is dominated by the exponentially growing component, hence the system is also unstable. 
\end{enumerate}
The above cases are listed in the Table.~\ref{tabS:stability cases} as well.

\subsection{Nonlinear effect for a 2-unit system}
As we mentioned in the Methodology, our robotic units actually follow a nonlinear force function (Eq.~(M3)). For a 2-unit system, the explicit constitutive relation reads as
\begin{equation}
\label{eqS:2unitsystem_realtorque}
    \binom{\tau_{1}}{\tau_{2}}=
    f\left(-
    \begin{bmatrix}
    k_{1}^{o}   & k_{1}^{p}-k_{1}^{a} \\ 
    k_{1}^{p}+k_{1}^{a} & k_{2}^{o}
    \end{bmatrix}
    \binom{\delta\theta_{1}}{\delta\theta_{2}}\right)
    -\begin{bmatrix}
    k^{e} & 0 \\ 
    0 & k^{e}
    \end{bmatrix}
    \binom{\delta\theta_{1}}{\delta\theta_{2}},
\end{equation}
The first term corresponds to the motor torque, constrained by its finite maximum value $\tau_{\text{max}}$. $f(x)$ is a bi-linear function that $f(x)=x,\ \text{if } |x|<\tau_{\text{max}}$ and $f(x)=\tau_{\text{max}},\ \text{if } |x|\ge\tau_{\text{max}}$. The second term is the restoring torque provided by the passive elastic skeleton. 

Taking into account this nonlinearity, the system effectively behaves as if subjected to a double-well potential and remains stable even when the motor torque saturates at $\tau_{\text{max}}$. This stabilization arises from the balance between the limited maximum torque that the motors can apply and the restoring torque from the elastic skeleton. We therefore revisit the stability of the aforementioned five cases and show the corresponding force fields of the nonlinear system. In addition, we conduct overdamped simulations and plot the resulting trajectories under a single external sine driving force. Specifically,
\begin{enumerate}
    \item When $\mathrm{Tr}(K)<0$ and $\Delta<0$ (the red regime in Extended Data Fig.~5a), the system becomes quadstable and the origin of the force field is a spiral source. Under a sine driving force applied to unit 1, the system initially settles into one stable fixed point. The trajectory sequentially traverses all four stable fixed points (Extended Data Fig.~5b(i)).
    \item When $\mathrm{Tr}(K)>0$ and $\Delta<0$, the system remains monostable and the origin is a spiral sink. The trajectory exhibits a trivial oscillatory motion (Extended Data Fig.~5b(ii)).
    \item When $\mathrm{Tr}(K)<0$, $\mathrm{det}(K)>0$ and $\Delta>0$, the system becomes quadstable and the origin of the force field is a source. The system also settles into one stable fixed point initially. The trajectory, however, only traverses two of the four stable fixed points (Extended Data Fig.~5b(iii)). 
    \item When $\mathrm{Tr}(K)>0$, $\mathrm{det}(K)>0$ and $\Delta>0$, the system remains monostable and the origin is a sink. The trajectory exhibits a trivial oscillatory motion (Extended Data Fig.~5b(iv)).
    \item When $\mathrm{det}(K)<0$, the system becomes bistable and the origin is a saddle point. The eigenvector corresponding to the negative real eigenvalue gives the direction of the unstable manifold. Under external driving, the system can switch between the two stable fixed points (Extended Data Fig.~5b(v)).
\end{enumerate}
These results are summarized in Table~\ref{tabS:stability cases}. Crucially these regimes encircle a critical exceptional point, the white dot in the middle of Extended Data Fig.~5a. It separates the monostable and multistable phases, at which point the eigenvalues and eigenvectors coalesce. We have explored it systemically in~\cite{al-izziNonreciprocalBucklingMakes2025}.

For a linearly stable system, introducing nonlinearity does not change its stability. However, for a linearly unstable system, multiple stable fixed points, i.e., stable shape changes, are generated while considering nonlinearity. As aforementioned, this is due to the balance between the limited maximum torque that the motors can apply and the restoring torque from the elastic skeleton. The magnitude of these stable fixed points is $\tau_{\text{max}}/k^{e}$. The number of stable fixed points is twice the number of eigenvalues with a negative real part of $K$. In the bistable case (Extended Data Fig.~5b(v)), there is one negative eigenvalue that leads to two unstable manifolds and gives two stable fixed points. Similarly, in the quadstable case (Extended Data Fig.~5b(i, iii)), there are two eigenvalues with a negative real part that lead to a source origin. The system will be stabilized at maximum torque along the directions of the two eigenvectors, so that there are four stable fixed points.

Another intriguing discovery is that adding non-reciprocity curls the force field of a quadstable reciprocity system (Extended Data Fig.~5b(iii)) and creates a unidirectional transition between different stable fixed points. We leverage it to achieve cyclic shape changing with a single driving signal in our metamaterials, as shown in Fig.~3g.

\begin{table}[htbp!]
    \caption{\rv{
    \textbf{The stability of a 2-unit system.}}}
    \label{tabS:stability cases}
    \centering
    \begin{tabular}{cccccc}
    \toprule\toprule
    \textbf{Tr($K$)} & \textbf{det($K$)} & \boldmath$\Delta$ & 
    \textbf{Origin} & \textbf{Linear system} & \textbf{Nonlinear system} \\ 
    \midrule
    -- & + & -- & spiral source & unstable & quadstable \\
    + & + & -- & spiral sink & stable & monostable \\
    -- & + & + & source & unstable & quadstable \\
    + & + & + & sink & stable & monostable \\
      & -- &   & saddle & unstable & bistable \\
    \bottomrule
    \end{tabular}
\end{table}

\subsection{Stability of an $N$-unit system}
For an $N$-unit system, the analytical solution of its linear overdamped dynamics is
\begin{equation}
    \delta\Theta(t)=\sum_{i=1}^{N}A_{i}\mathbf{V}_{i}e^{-\lambda_{i}t}.
\end{equation}

The stability of an $N$-unit system is determined by the eigenvalues $\lambda_{i}$ of the motor stiffness matrix $K$. If all eigenvalues are positive real numbers, the system is monostable. If there is at least one eigenvalue with a negative real part, the system exhibits multistability while considering the nonlinear motor saturation. The number of stable fixed points, i.e., stable shape changes, is twice the number of eigenvalues with a negative real part. The direction of unstable manifolds, i.e., how the system deforms, is determined by the eigenvectors corresponding to the eigenvalues with a negative real part. The magnitude of the stable fixed points is $\tau_{\text{max}}/k^{e}$, but we can vary $\tau_{\text{max}}$ digitally to determine the positions of stable fixed points. Therefore, to learn multistable shape changes in our metamaterials, the key is to trigger eigenvalues with a negative real part. We discuss this in the following section.
}

\section{Contrastive learning with stability constraints}
\label{SM_stability constraint}
As we mentioned above, the existence of eigenvalues with a negative real part leads to multistability in our metamaterials. But how to trigger or avoid eigenvalues with a negative real part locally so that we can control the system stability during contrastive learning? Here, we introduce a local stability constraint.

Our stability constraint rule is based on the Gershgorin circle theorem~\cite{semyon_aronovich_uber_1931}. For a square $n \times n$ matrix $\mathcal{A}$, the theorem states that each eigenvalue of $\mathcal{A}$ lies within at least one of the Gershgorin discs. These discs are set in a complex space. The center and radius of each Gershgorin disk are simply defined using the information from each row of $\mathcal{A}$. Let $D_{i}$ be the sum of the absolute values of the off-diagonal entries in the $i^{\textrm{th}}$ row as $D_{i}=\sum_{i\neq j}^{n}|a_{ij}|$. A Gershgorin disk $\mathcal{D}(a_{ii}, D_{i})$ is defined as a circle with a center of the diagonal entry $a_{ii}$ and a radius of $D_{i}$ in the complex space. 

Using the Gershgorin circle theorem, we impose a local constraint on the eigenvalues of the stiffness matrix $K$. Considering  Eq.~(1), $K$ is a tridiagonal matrix, we have that $D_{i}=|k_{i-1}^{p}+k_{i-1}^{a}|+|k_{i}^{p}-k_{i}^{a}|$ and $a_{ii}=k_{i}^{o}+k^{e}$. 

In order to ensure the system is monostable, we have to make sure there are no eigenvalues with a negative real part in $K$ according to our stability analysis. So the following stability constraint must be imposed during contrastive learning:
\begin{equation}
\label{eqS:constraint_mono} 
\begin{cases}
    k_{i}^{o}+k^{e}>0, & \forall i\\
    D_{i}<|k_{i}^{o}+k^{e}|, & \forall i. 
\end{cases}
\end{equation}
After each epoch, the stiffnesses stop evolving if any unit violates the above constraint. Eq.~\eqref{eqS:constraint_mono} makes sure the Gershgorin discs are located in the positive real part of the complex space so that all eigenvalues have positive real parts. 

In order to ensure the system is multistable, there should be at least one eigenvalue with a negative real part. So there is at least one unit $i$ for which 
\begin{equation}
\label{eqS:constraint_multi}
    \begin{cases}
        k_{i}^{o}+k^{e}<0, \\
        D_{i}<|k_{i}^{o}+k^{e}|.
    \end{cases}
\end{equation}
This constraint promises that one Gershgorin disk is located in the negative real part of the complex space. With this stability constraint, we can now trigger multistability during contrastive learning. \rv{The easiest way is just pushing the onsite stiffness $k_{i}^{o}$ smaller than $k^{e}$ so that $\mathrm{Tr}(K)$ is negative. To do this, we impose an extra gradient descent (Eq.~(8)) on a set of units $\mathcal{M}$ and thus push their on-site stiffness $k^{o}$ to be negative.} This ensures that the units $i$ in $\mathcal{M}$ follow the above stability constraint (Eq.~\eqref{eqS:constraint_multi}) so that eigenvalues with negative part appear during learning. We use this constrained learning rule to train multistable metamaterials and demonstrate robotic applications (Figs.~3d-g and Movie.~S4).

\rv{
\section{Simpler variants of the learning rule}
While the gradient-based contrastive learning used in our work demonstrates great learning performance, it comes with certain hardware constraints. Specifically, it requires high-resolution information, i.e, the actual angular deflections. To address this limitation, exploring simpler learning rules that do not require high-precision sensors and complex processors is also a valuable idea to extend. Inspired by \cite{sternSupervisedLearningPhysical2020a}, we propose a simplified variation of the contrastive learning rule and verify its feasibility in numerical simulations. We first show it works for learning simple shape changes, and then extend it to learning non-reciprocal cases. We also compare this binary rule to the learning rule used in the Main Text.

Here, instead of using onsite, symmetric and anti-symmetric stiffness components ($k_{i}^{o}$, $k_{i}^{p}$ and $k_{i}^{a}$) as in the Main Text, we use general entries $k_{i,j}$ as well as the learning degrees of freedom to build up the stiffness matrix.

\subsection{Binary contrastive learning rule}
The learning protocol is the same as that used in the Main Text. We binarize the contrastive learning rule (Eq.~(2)) and it only needs binary information to update stiffness. Specifically, only measuring if the angles of output in the clamped state are higher or lower than those in the free state, and the sign of each angle. This binary contrastive learning rule (Eq.~(2)) reads as
\begin{equation}
\label{eqS:binary contrastive learning rule}
\frac{\mathrm{d}k_{i,j}}{\mathrm{d}t}=
\begin{cases}
     -\gamma \,\mathrm{sgn}(\delta\theta_{i}^{C}-\delta\theta_{i}^{F})\mathrm{sgn}(\delta\theta_{j}^{F}), & \text{for } |i-j|=1, \\
     \dfrac{\gamma}{2}\,\mathrm{sgn}(\delta\theta_{i}^{C}-\delta\theta_{i}^{F})\mathrm{sgn}(\delta\theta_{j}^{F}), & \text{for } i=j.
\end{cases}
\end{equation}
Here, the first equation is for updating the off-diagonal terms, i.e., the interactions between the unit $i$ and $j$. $\mathrm{sgn}(\delta\theta_{i}^{C}-\delta\theta_{i}^{F})$ shows whether the angular deflection of the $i^{\text{th}}$ unit in the free state is higher (lower) than that in the clamped state. $\mathrm{sgn}(\delta\theta_{j}^{F})$ tells the direction of torque contribution from the neighbor unit $j$. Their product ensures that stiffness $k_{i,j}$ is adapted in the direction that reduces the mismatch between free and clamped states. The second equation is for updating the diagonal term, i.e., the onsite stiffness of the unit $i$. The factor $\frac{1}{2}$ keeps it consistent with Eq.~(4) in the Main Text. Similarly, the product means that the onsite stiffness $k_{i,i}$ should be reduced or increased depends on whether $\delta\theta_{i}^{F}$ is above or below $\delta\theta_{i}^{C}$ and $\delta\theta_{i}^{F}$ is positive or negative. This rule is much simpler than the one used in the Main Text because it only needs binary measurements, i.e., whether $\delta\theta_{i}^{F}$ is higher or lower $\delta\theta_{i}^{C}$, and check if $\delta\theta_{i}^{F}$ is positive or negative.

We use this binary learning rule and let a system with $N=6$ learn the same task in Fig.~1b, morphing into a U-shape. In Fig.~\ref{figS:letter_U_Binary}a, our metamaterial is also able to learn to form a U-shape by using this simple binary learning rule (Eq.~\eqref{eqS:binary contrastive learning rule}). We compare its error curve to that of complete contrastive learning (Eq.~(2)) and find that the system equipped with the binary learning rule learns more slowly. It is not surprising because it follows the shortest path that decreases the local error and ignores the gradient magnitude. In addition, the MSE of (Eq.~\eqref{eqS:binary contrastive learning rule}) oscillates at around $10^{-5}$ after ~200 epochs. This is because the binary learning method cannot reach the optimum, and it keeps pushing the system slightly above and below the optimal solution. In addition, there is a small jump at around 290 epochs for the binary learning rule (Eq.~\eqref{eqS:binary contrastive learning rule}). It arises from $k_{1,1}$ becoming negative, thereby changing the stability of the system. However, this binary learning rule can still find another optimal solution afterward. 

We notice that the stiffness matrix of the binary learning rule evolves to be asymmetric eventually (Fig.~\ref{figS:letter_U_Binary}b). Since our binary rule is not explicitly symmetrized, this asymmetry arises simply from random noise filling the off-diagonal entries. Next, we test if (Eq.~\eqref{eqS:binary contrastive learning rule}) can also learn non-reciprocal shape changes as depicted in Fig.~2a. Specifically, applying a positive curvature to the left part leads to a positive curvature to the right part, whereas applying a positive curvature to the right part leads to a negative curvature to the left part. In Fig.~\ref{figS:letter_US_Binary}a, the error shows that this binary learning rule fails to learn non-reciprocal shape changes. Therefore, we continue to explore and see if we can adapt this binary learning rule to learn non-reciprocal shape changes.

\begin{figure}[htbp!]
    \centering
    \includegraphics[width=1\linewidth]{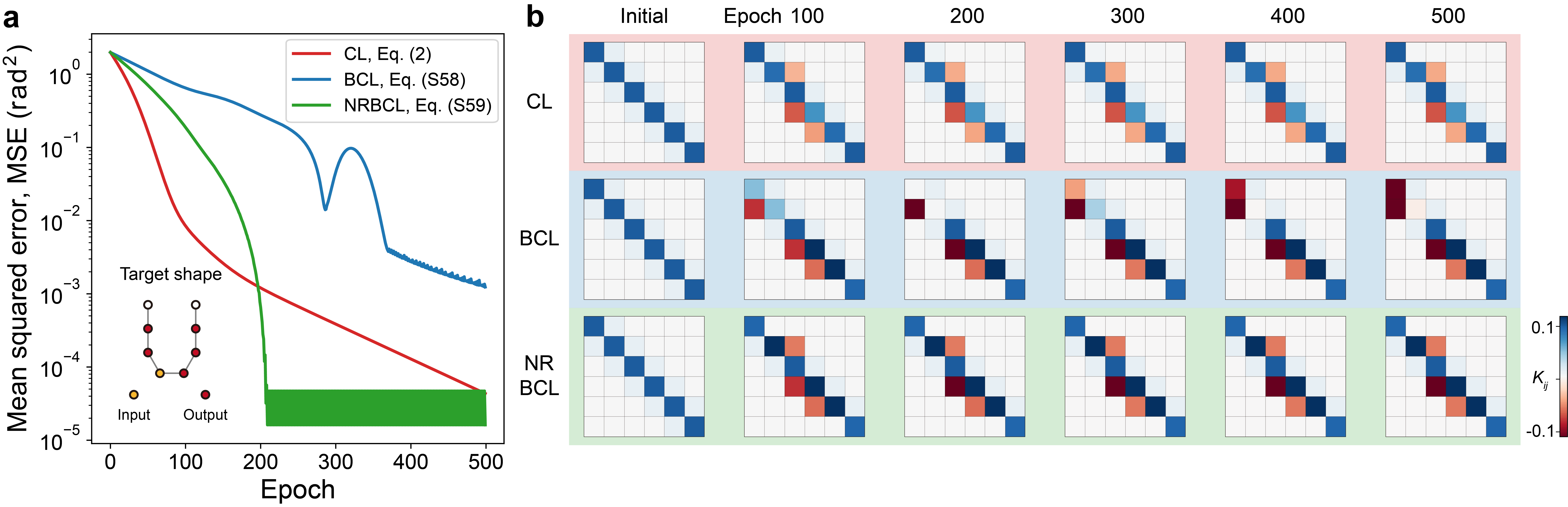}
    \caption{\rv{
    \textbf{Learning to morph a U-shape by using different variants of the learning rule.} 
    \textbf{a,} The MSE curves of learning a U-shape by using the complete contrastive learning rule (CL, Eq.~(2)), the binary contrastive learning rule (BCL, Eq.~\eqref{eqS:binary contrastive learning rule}) and the non-reciprocal binary contrastive learning rule (NRBCL, Eq.~\eqref{eqS:NR binary contrastive learning rule}). 
    \textbf{b,} The corresponding stiffness matrices $K$ during training. Here, we use the system with $N=6$ units and with the first nearest-neighbor interaction. The initial parameters for CL are $k_{i}^{o}=0.1$, $k_{i}^{p}=0.01$ and $k_{i}^{a}=0.0$ and those of BCL and NRBCL are $k_{i,i}=0.1$ and $k_{i,j}=0.01$ for $|i-j|=1$. The learning rate $\gamma$ is 0.01.}
    }
    \label{figS:letter_U_Binary}
\end{figure}

\subsection{Path-dependent binary contrastive learning rule}
In order to extend (Eq.~\eqref{eqS:binary contrastive learning rule}) to learning non-reciprocal shape changes as well, we modify the learning rule by adding a path-dependent term to
\begin{equation}
\label{eqS:NR binary contrastive learning rule}
\frac{\mathrm{d}k_{i,j}}{\mathrm{d}t}=
\begin{cases}
     -\gamma \,\mathrm{sgn}(\delta\theta_{i}^{C}-\delta\theta_{i}^{F})\mathrm{sgn}(\delta\theta_{j}^{F}), & \text{for } i-j=\alpha_{i}, \\
     \dfrac{\gamma}{2}\,\mathrm{sgn}(\delta\theta_{i}^{C}-\delta\theta_{i}^{F})\mathrm{sgn}(\delta\theta_{j}^{F}), & \text{for } i=j,
\end{cases}
\end{equation}
where $\alpha_i=\mathrm{sgn}(i-I)$ for $i\neq I$, or $\alpha_i=\mathrm{sgn}(O-I)$ for $i=I$. The path-dependent term $\alpha$ here is similar to $\alpha$ in the Main Text: it also indicates the loading path between unit $i$ and input units $I$. If the $i^\textrm{th}$ unit is on the right side of the input $I$ ($i > I$), the loading path goes from left to right, $\alpha_i=1$, and only the lower off-diagonal entries in the stiffness matrix $K$ are updated. In contrast, if the $i^\textrm{th}$ unit is on the left side of the input $I$ ($i < I$), the loading path goes backward from right to left, $\alpha_i=-1,$ and only the upper off-diagonal entries in the stiffness matrix $K$ are updated. 

We now use the modified learning rule and let the same system learn the non-reciprocal shape changes depicted in Fig.~2a. In Fig.~\ref{figS:letter_US_Binary}a, we show the MSE. The modified learning rule (Eq.~\eqref{eqS:NR binary contrastive learning rule}) learns to generate the non-reciprocal shape changes successfully. Surprisingly, it learns more quickly than the complete contrastive learning rule. We conjecture that this is because binary learning traverses the error landscape along a more direct path than gradient descent by ignoring the true gradient of the error landscape; it only cares about the sign of the error and essentially performs a greedy algorithm. This direct path sometimes performs better than the gradient descent path, which is used in the complete contrastive learning method (Eq.~(2)).

These findings suggest that simplified learning rules—requiring only binary local information—can also achieve adaptive behaviors. It highlights that physical learning in our mechanical system does not necessarily require full information like the loss function and gradient, as well as high-precision sensors and complex processors, which is encouraging for future hardware-constrained implementations. 
}

\begin{figure}[htbp!]
    \centering
    \includegraphics[width=1\linewidth]{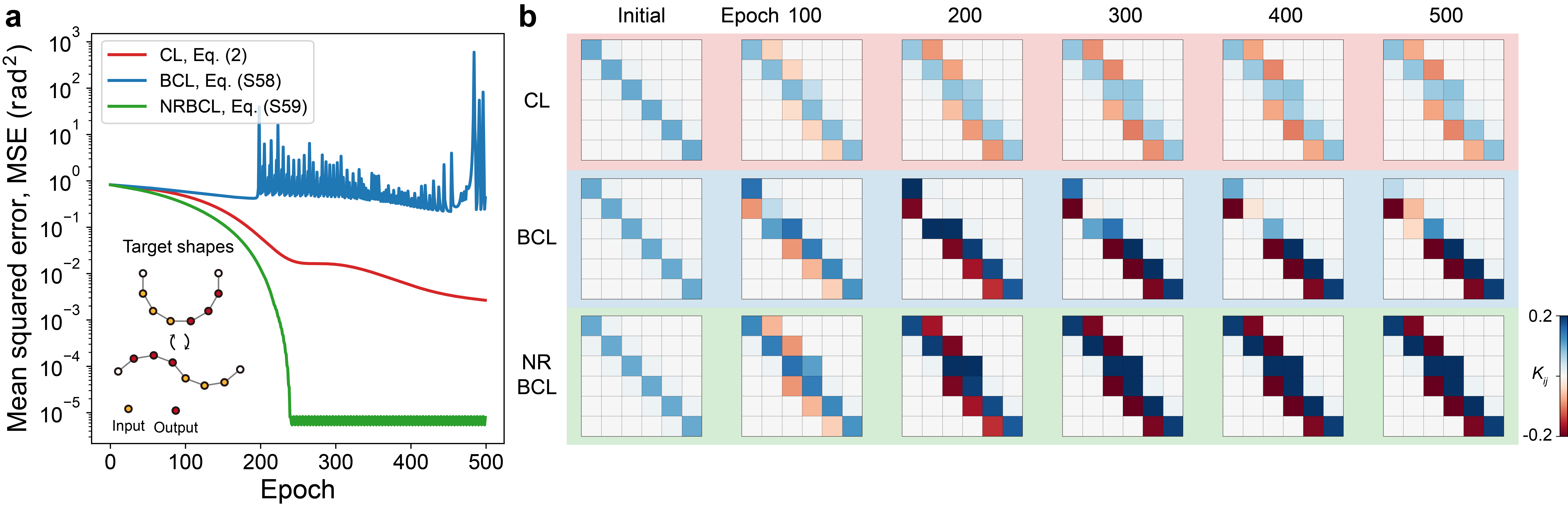}
    \caption{\rv{
    \textbf{Learning non-reciprocal shape changes by using different variants of the learning rule.} 
    \textbf{a,} The MSE curves of learning a U-shape by using the complete CL rule, (Eq.~(2)), the BCL rule, (Eq.~\eqref{eqS:binary contrastive learning rule}), and the NRBCL rule (Eq.~\eqref{eqS:NR binary contrastive learning rule}). 
    \textbf{b,} The corresponding stiffness matrices $K$ during training. The simulation protocol is the same as that in Fig.~\ref{figS:letter_US_Binary}.}
    }
    \label{figS:letter_US_Binary}
\end{figure}

\end{document}